\documentclass[10pt]{article}

\usepackage[english]{babel}
\usepackage[utf8]{inputenc}
\usepackage[T1]{fontenc}
\usepackage{gensymb}
\usepackage{float}
\usepackage{algorithm}
\usepackage[noend]{algpseudocode}
\usepackage[margin=0.75in]{geometry}
\usepackage[backend=bibtex, sorting=none]{biblatex} 
\usepackage{amsmath}
\usepackage{enumitem}
\usepackage{graphicx}
\usepackage[colorinlistoftodos]{todonotes}
\usepackage{hyperref}
\usepackage[title]{appendix}
\usepackage{multirow}
\usepackage{subfig}
\usepackage{authblk}
\usepackage{multicol}
\usepackage{titlesec}
\usepackage{amssymb}

\titlespacing\section{0pt}{0pt plus 2pt minus 2pt}{0pt plus 2pt minus 2pt}
\title{Charged particle tracking \\ with quantum annealing-inspired optimization}

\author[1]{Alexander Zlokapa}
\author[2]{Abhishek Anand}
\author[1]{Jean-Roch Vlimant}
\author[3,6]{Javier M. Duarte}
\author[4]{Joshua Job}
\author[5]{Daniel Lidar}
\author[1]{Maria Spiropulu}

\affil[1]{Division of Physics, Mathematics \& Astronomy, Alliance for Quantum Technologies, California Institute of Technology, Pasadena, CA 91125, United States}
\affil[2]{Harvard University, Cambridge, MA 02138, United States}
\affil[3]{Fermi National Accelerator Laboratory, Batavia, IL 60510, United States}
\affil[4]{Lockheed Martin Corporation, El Segundo, CA 90245, USA}
\affil[5]{Departments of Electrical and Computer Engineering, Chemistry, and Physics \& Astronomy, and Center for Quantum  Information Science \& Technology, University of Southern California, Los Angeles, CA 90089, United States}
\affil[6]{University of California San Diego, La Jolla, CA 92093, United States}

\date{\vspace{-5ex}}

\begin{document}

\maketitle

\begin{abstract}
\normalsize
At the High Luminosity Large Hadron Collider (HL-LHC), traditional track reconstruction techniques that are critical for analysis are expected to face challenges due to scaling with track density. 
Quantum annealing has shown promise in its ability to solve combinatorial optimization problems amidst an ongoing effort to establish evidence of a quantum speedup. As a step towards exploiting such potential speedup, we investigate a track reconstruction approach by adapting the existing geometric Denby-Peterson (Hopfield) network method to the quantum annealing framework and to HL-LHC conditions. Furthermore, we develop additional techniques to embed the problem onto existing and near-term quantum annealing hardware. Results using simulated annealing and quantum annealing with the D-Wave 2X system on the {\it TrackML} dataset are presented, demonstrating the successful application of a quantum annealing-inspired algorithm to the track reconstruction challenge. We find that combinatorial optimization problems can effectively reconstruct tracks, suggesting possible applications for fast hardware-specific implementations at the LHC while leaving open the possibility of a quantum speedup for tracking.
\end{abstract}
\newpage
\begin{multicols}{2}
\section{Introduction}
Track reconstruction is a critical and computationally intensive step for data analysis at high energy particle accelerator experiments \cite{albrecht2019roadmap}. The High-Luminosity LHC (HL-LHC) upgrade, which is expected to be completed in 2026, will increase the number of simultaneous collisions (pileup) per proton bunch crossing  from approximately 40 to up to 200~\cite{ApollinariG.:2017ojx}. 
Under these conditions, conventional algorithms such as a Kalman filter scale worse than quadratically with respect to the number of hits and are thus expected to require excessive computing resources~\cite{albrecht2019roadmap}. A variety of alternatives to current particle tracking methods are being pursued~\cite{Cerati_2018,cachep2014, Farrell:2018cjr} to tackle the enhanced combinatorics of tracking at the HL-LHC.

It is an open question as to whether quantum annealing (QA) implementable in current hardware offers any scaling speedup over classical methods. However, for specific optimization problems, quantum annealing~\cite{PhysRevE.58.5355} outperforms classical heuristics like simulated annealing (SA)~\cite{farhi2002quantum, PhysRevX.8.031016}, and competitive performance has already been demonstrated for certain machine learning tasks~\cite{Mott:2017aa,Li:comp-bio-2017}. It thus may present a promising avenue for tracking if we can represent it as an appropriate optimization problem. We describe here a prototype for a charged particle track reconstruction method using a programmable quantum annealer.

We first provide background information for both track reconstruction and QA in section~\ref{sec:bkg}. In section~\ref{sec:hllhc-db}, we formulate the tracking problem in the annealing framework, describe the dataset and the algorithms required to solve the problem with quantum annealing. We present the results of our method, and discuss its algorithmic complexity in section~\ref{sec:results}. We provide a concluding analysis of potential real-world applications in section~\ref{sec:related}. section~\ref{sec:related}.

\section{Background}
\label{sec:bkg}
\subsection{Track Reconstruction}
Reconstructing the path of a charged particle produced in a collider like the LHC is an essential task for the analysis of the collider experiment's data. The track of a charged particle within a magnetic field is locally approximated by a helix. Measurement of the curvature of this helix enables the determination of the components of the particle's momentum that are transverse to the magnetic field.
Furthermore, in collider physics, tracks are crucial for a variety of measurements such as reconstruction of decay vertices~\cite{Collaboration_2014}, identification of jet flavor~\cite{cmsrun1btag,cmsrun2btag,atlasrun1btag,atlasrun2btag}, pile-up mitigation~\cite{CMS-PAS-JME-14-001,cmsrun1met,cmsrun2met} and are particularly important in complementing calorimeter measurements at low energy.
Tracks are also necessary for global event description algorithms, such as the particle flow reconstruction algorithm~\cite{cmspf}, which can improve the accuracy and resolution of many key observables in particle physics. 

One of the key elements of data acquisition at a hadron collider experiment is the trigger system~\cite{Khachatryan_2017} that selects, in almost real time, the most interesting collision events from the rudimentary high rate events, which number many orders of magnitude more. The trigger system reduces the rate of collisions under consideration from 40~MHz to 1--2~kHz with fast algorithms, dedicated on-board on-chip hardware and subsequent processing in software farms. In this context, the reconstruction of the charged particles must be done quickly (on the order of $\mu$s at the hardware level) and efficiently. The approach proposed in this work may eventually provide a solution for fast tracking in the trigger.

\begin{figure}[H]
\centering
\includegraphics[width=0.35\textwidth]{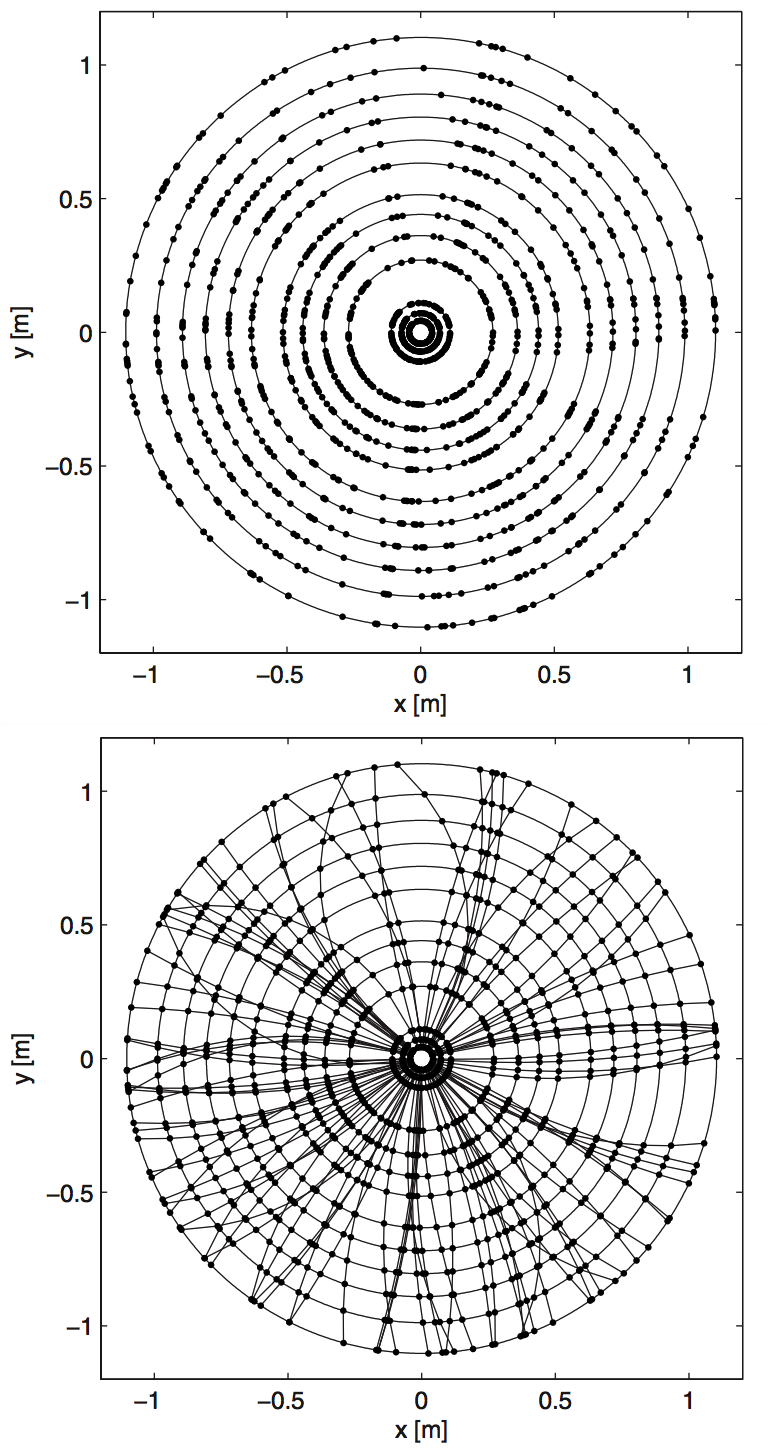}
\caption{Display, in the plane transverse to the beam, of the hits left in the tracker (top). The cylindrical geometry is typical of collider tracker devices. Display of the simulated tracks and the hits they left in the tracker (bottom). The goal of charged particle tracking is to cluster the hits in the original trajectories ~\cite{strandlie2010track}.
}
\label{fig1}
\end{figure}

\begin{figure}[H]
\centering
\includegraphics[width=0.35\textwidth]{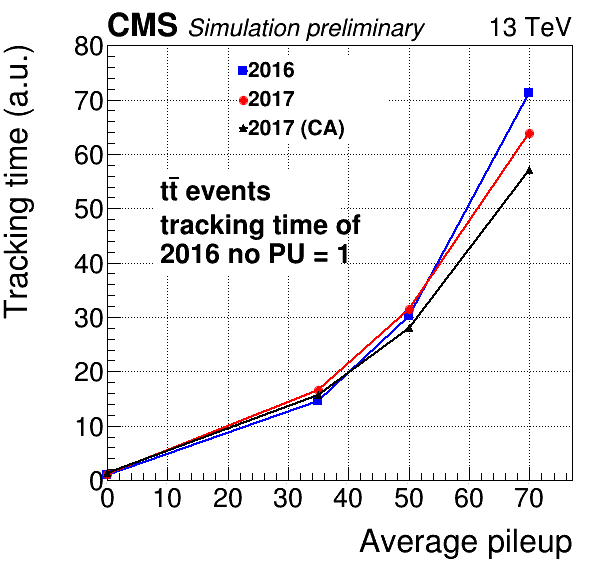}
\caption{Timing performance of 2016 and 2017 CMS conventional tracking methods, as well as 2017 tracking with cellular automaton (CA) seeding \cite{trackingscale} with respect to average pileup. Note that currently LHC collisions contain an average of 40 pileup interactions corresponding to a multiplicity of 2,000 particles (tracks). HL-LHC is expected to operate at an average of 200 pileup.}
\label{fig:scalefig}
\end{figure}

The problem of track reconstruction (figure~\ref{fig1}) can be formally stated as follows: given a set of hits (detector-particle interactions) with different spatial positions, the goal is to cluster hits into collections of hits that are believed to come from the same particle.
The current methods used for tracking can be broadly classified into sequential and global methods.  Sequential methods construct tracks one by one: for example, the road method~\cite{strandlie2010track} and the Kalman filter~\cite{billoir1984track}. Global methods construct all tracks at once and are, at the core, clustering algorithms in some feature space: for example, the Hough transform~\cite{hough1959machine, CHESHKOV200635} and Hopfield network also called the Denby-Peterson network~\cite{denby1988neural, peterson1989track, stimpfl1991fast}. 
Most of these methods scale worse than quadratically with the number of tracks per event.
In particular, the scaling of the combinatorial track finder algorithm (see figure~\ref{fig:scalefig}) as a function of the number of concurrent proton-proton interaction per bunch crossing in the LHC (so called pileup) would no longer be feasible at higher track density.

\subsection{Quantum annealing}
Quantum annealing is a form of adiabatic quantum computing (AQC)~\cite{PhysRevX.8.031016}. Informally, the adiabatic theorem states that if a quantum mechanical system starts in the ground state of some Hamiltonian, and we change the Hamiltonian slowly enough, the system will end in the ground state of the final Hamiltonian. This theorem is exploited for computation by setting the initial Hamiltonian to be one with a known ground state and setting the final Hamiltonian to be the problem Hamiltonian whose ground state represents the solution to the optimization problem we wish to solve. The annealing time scale (expected time needed to reach the solution in a single run) is bounded by the inverse of the smallest energy gap between the ground and first excited state, encountered during the adiabatic evolution~\cite{PhysRevX.8.031016}. 

Using this scheme, we can solve certain combinatorial optimization problems~\cite{farhi_quantum_2000}. Specifically, it has been used to solve problems where the problem Hamiltonian represents the Hamiltonian of an Ising spin system (a mathematical model of magnetism in statistical mechanics). More generally, any quadratic unconstrained binary optimization (QUBO) problem can be naturally mapped to an Ising spin problem and can be encoded into the machine Hamiltonian~\cite{2013arXiv1302.5843L}. QUBO problems can be formally expressed as: 
\begin{equation}
\min_\mathbf{X} E(\mathbf{X})=\sum _{i}^{N}h_{i}X_{i}+\sum _{i<j}^{N} J_{ij}X_{i}X_{j} ,
\end{equation}
where $X_i \in \{0, 1\}$ are the components of $\mathbf{X}$, $h_i \in \mathbb{R}$ represents an external interaction, and $J_{ij}\in \mathbb{R}$ represents a two-body interaction. The objective is to find the assignment of $\mathbf{X}$ that minimizes $E$. This becomes the Hamiltonian of an Ising model after replacing each $X_i$ by $\frac{1}{2}(s_i+1)$, where $s_i \in \{-1, 1\}$, and dropping the resulting constant term $\frac{1}{2}\sum_i h_i + \frac{1}{4}\sum_{i<j} J_{ij}$.

\section{Track reconstruction as a QUBO}
\label{sec:hllhc-db}
We map the track reconstruction problem to a QUBO problem through a procedure inspired by the Denby-Peterson method~\cite{denby1988neural, peterson1989track}. However, we make modifications to improve its performance for the HL-LHC, adding specific terms to the QUBO that correspond to LHC-type detector geometry and conditions. Finally, we present classical pre-processing heuristics that are computationally efficient, allowing us to evaluate the track reconstruction problem on a programmable quantum annealer and using SA.
\subsection{Denby-Peterson method}
\label{sec:dp}
The Denby-Peterson (DP) track reconstruction method~\cite{denby1988neural, peterson1989track} interprets the track reconstruction problem as an track segment classification problem. It has been earlier proposed and validated for the ALEPH~\cite{stimpfl1991fast}, ARES~\cite{baginyan1994tracking}, and ALICE~\cite{pulvirenti2004neural} experiments with encouraging results. More recently, the method has been deployed in the LHCb experiment muon system~\cite{passaleva2008recurrent}.

\begin{figure}[H]
\centering
\includegraphics[width=0.35\textwidth]{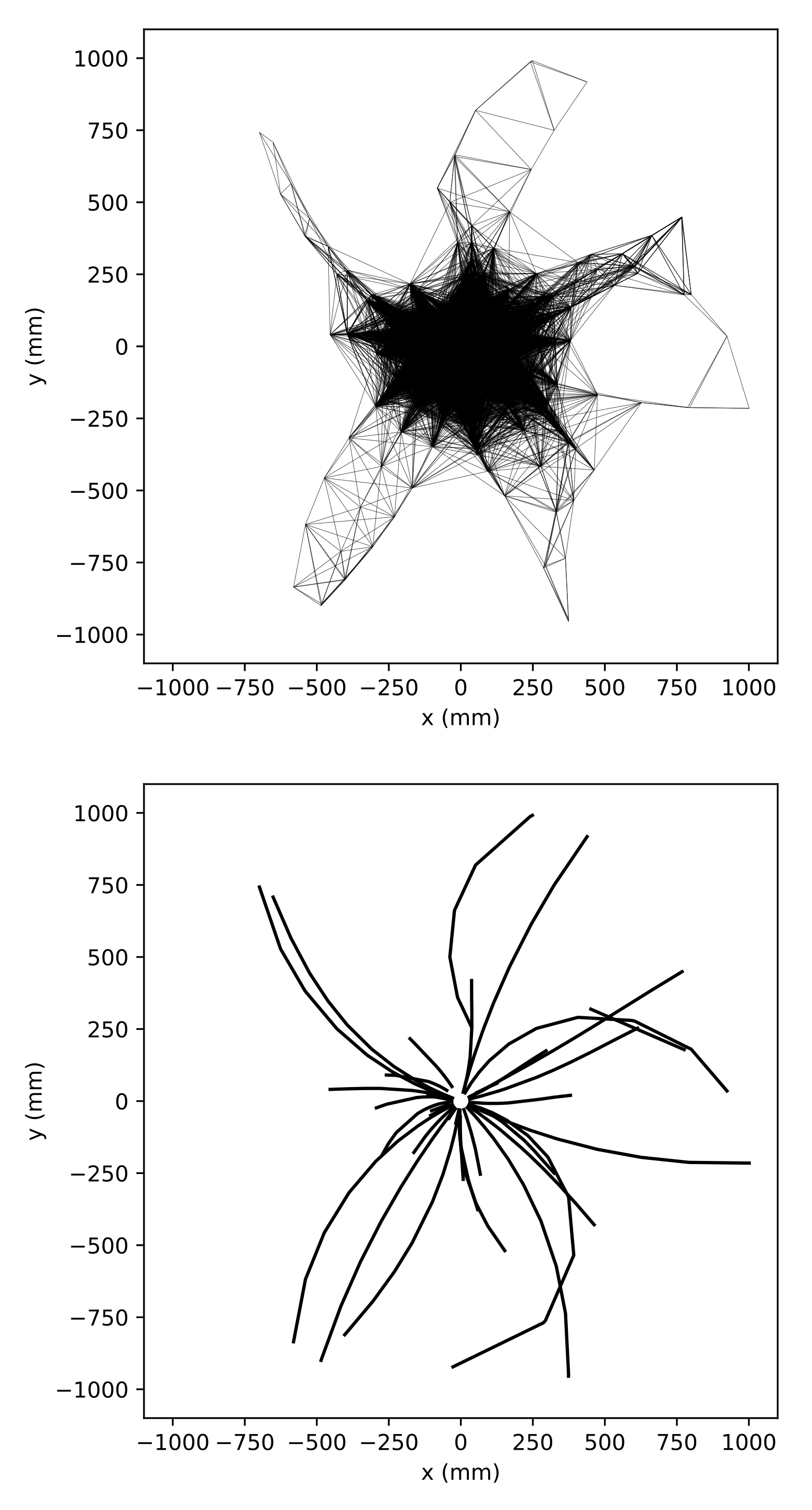}
\caption{Projection in the transverse plane of 50 tracks from one event in the TrackML dataset. In the Denby-Peterson algorithms, all pre-selected potential edges are considered (top), and only the relevant one remain after optimization (bottom).}
\label{fig:xy}
\end{figure}

The DP method optimizes an energy function that resembles a QUBO. Before briefly describing the original algorithm, we schematically show its intended results in figure~\ref{fig:xy}. We begin with edges between pre-selected pairs of hits (top of figure~\ref{fig:xy}). After optimizing the DP energy function, we expect a value of 1 to be assigned to all correct edges (bottom of figure~\ref{fig:xy}) and a value of 0 to be assigned to all incorrect edges.

Let the set $S$ contain $N$ binary variables $s_{ab}$ representing all unique edges between a hit $a$ and a hit $b$ subject to the constraint that hit $a$ is closer to the center than hit $b$ (for uniqueness). If $s_{ab}=1$ then the two hits are assumed to have been created by the same particle. The total energy of any given set $S$ is then given by~\cite{denby1988neural, peterson1989track, stimpfl1991fast}:
\begin{align}
\begin{split}
E = -\frac{1}{2}\Bigg[&\sum_{a, b, c} \left(\frac{\cos^\lambda{\theta_{abc}}}{r_{ab} + r_{bc}}s_{ab}s_{bc}\right) \\
&- \alpha\left(\sum_{b\neq c} s_{ab} s_{ac} + \sum_{a\neq c} s_{ab} s_{cb} \right)\\
&- \beta\left(\sum_{a, b} s_{ab} - N\right)^2\Bigg] ,
\end{split}
\end{align}
where  $\theta_{abc}$ is the Cartesian angle between the two line segments when they are transformed into cylindrical coordinates (i.e. helical tracks appear as straight lines as in figure~\ref{fig:rz}), $r_{ab}, r_{bc}$ are the line segment lengths, and $\lambda$  free parameter that help distinguish similar angles.
Each term corresponds to geometric rewards and penalties weighted by parameters $\alpha$ and $\beta$, biasing the tracks to be composed of short track segments that lie on a smooth curves with no bifurcations. 

Although the DP method offers a good starting point for tracking, several modifications can be made to the QUBO to provide it with additional information describing the HL-LHC configuration. In particular, we may encode expectations of the particles' trajectories and the detector geometry to simplify the optimization problem, enabling larger events to be successfully annealed. 

\subsection{Modified QUBO for HL-LHC} \label{sec:modified-qubo}
\begin{figure}[H]
\centering
\includegraphics[width=1.5in]{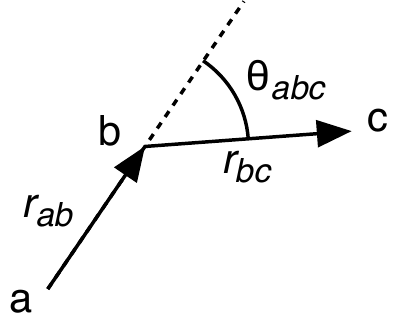}
\caption{Representation of three hits ($a,b,c$), the segments ($r_{ab}$, $r_{bc}$) and the opening angle in cylindrical coordinate $\theta_{abc}$. The angle $\phi_{abc}$ (not represented) is measured in the transverse plane.}
\label{fig3}
\end{figure}

\begin{figure}[H]
\centering
\includegraphics[width = 0.45\textwidth]{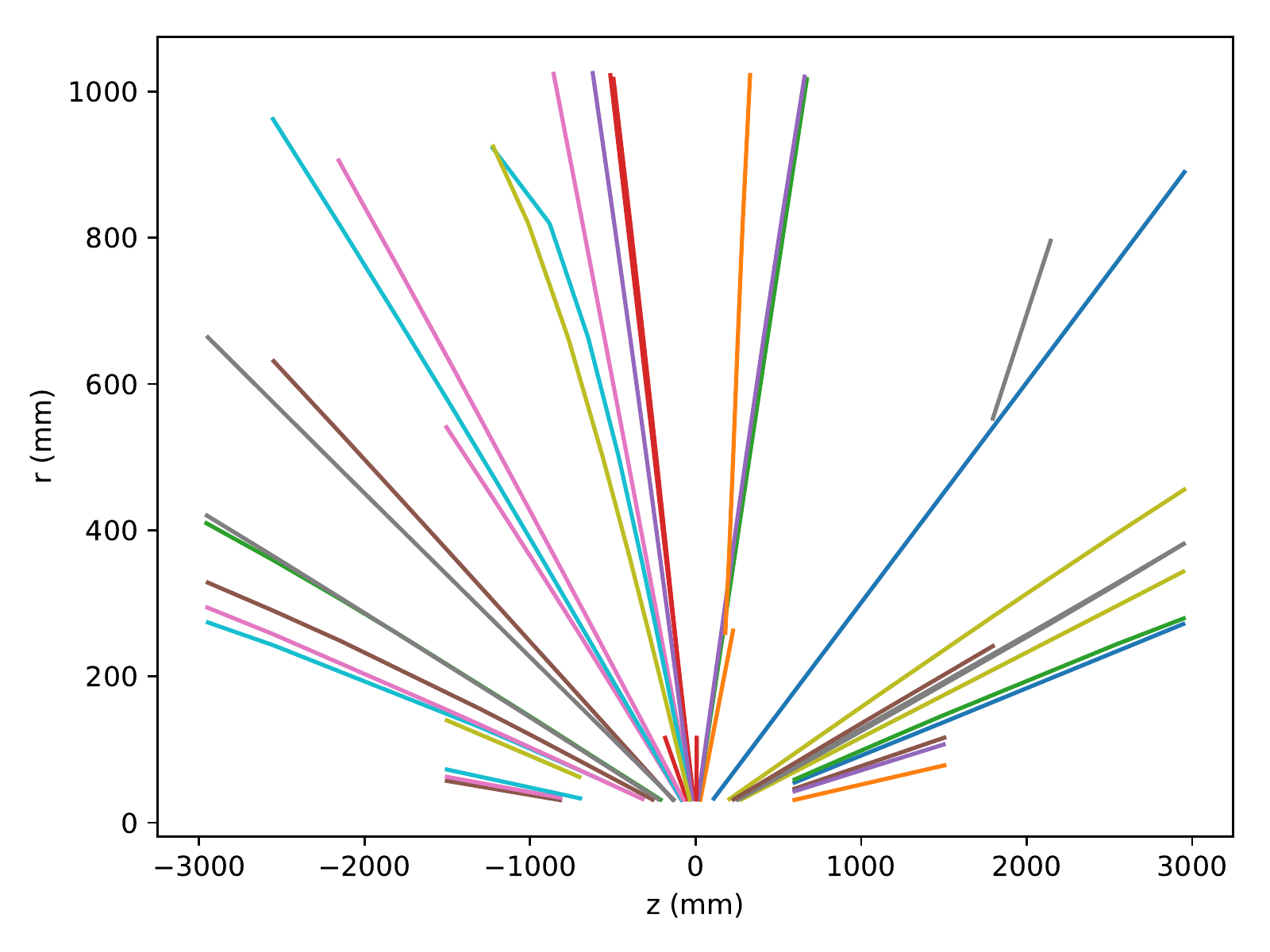}
\caption{Projection of 50 tracks in the $rz$-plane. Due to the uniform magnetic field, charged particles travel in straight lines in cylindrical coordinates. Hence, we may bias the QUBO towards perfectly helical tracks through the $\theta_{abc}$ term.}
\label{fig:rz}
\end{figure}

We begin with the same QUBO formulation:
\begin{align}
\begin{split}
E = -\frac{1}{2}\Bigg[&\sum_{a,b}\left(W^{\mathrm{reward}}_{ab} - W^{\mathrm{penalty}}_{ab}\right) s_{ab} \\
&+ \sum_{a,b,c}\left(U^{\mathrm{reward}}_{abc} - U^{\mathrm{penalty}}_{abc}\right) s_{ab}s_{bc}\Bigg].
\end{split}
\end{align}
We define the geometric reward to match the helical tracks observed in the LHC. If segments $s_{ab}$ and $s_{bc}$ share a point $b$, the reward is given by (see figure \ref{fig3}):
\begin{equation}
\frac{\cos^\lambda(\theta_{abc}) + \rho \cos^\lambda(\phi_{abc})}{r_{ab} + r_{bc}} ,
\end{equation}
where $\phi_{abc}$ is the azimuthal angle between the line segments in rectangular coordinates (i.e., helical tracks appear helical as in figure~\ref{fig:xy}).
Since we wish to track charged particles moving in a uniform magnetic field, we expect them to trace helical paths. The $\theta_{abc}$ term models this expectation, while the $\phi_{abc}$ term biases tracking towards high-momentum particles with weight $\rho$. By dividing by track length, we bias the tracking algorithm to favor a chain of short track segments. Note that we standardize the space in $r, \phi, z$ by dividing by characteristic lengths $1000, \pi$ and $1000$ respectively so the tracking is not biased in any particular coordinate direction. Furthermore, we threshold the edge affinity term to encourage a sparse graph for annealing, setting the reward weight to 0 if $\cos^\lambda(\theta_{abc}) < \tau$ for a free parameter threshold $\tau = 0.996$.

\begin{figure}[H]
\centering
\includegraphics[width = 0.3\textwidth]{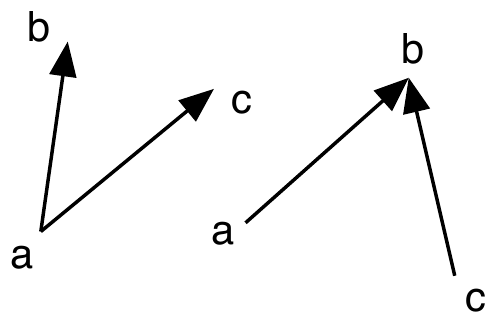}
\caption{
Representation of the segment configuration that are penalized with the term in equation~\ref{equ:bifurcation}.
}
\label{fig:bifurcation}
\end{figure}

As in the original DP method, we add a penalty for bifurcation: 
\begin{equation}\label{equ:bifurcation}
    \sum_{b\neq c} s_{ab} s_{ac} + \sum_{a\neq c} s_{ab} s_{cb}
\end{equation}
The two sums over $b\neq c$ and $a \neq c$ correspond to an inhibition of sharing hits at the beginning and the end of the segment respectively (see figure~\ref{fig:bifurcation}).

\begin{figure}[H]
\centering
\includegraphics[width = 0.5\textwidth]{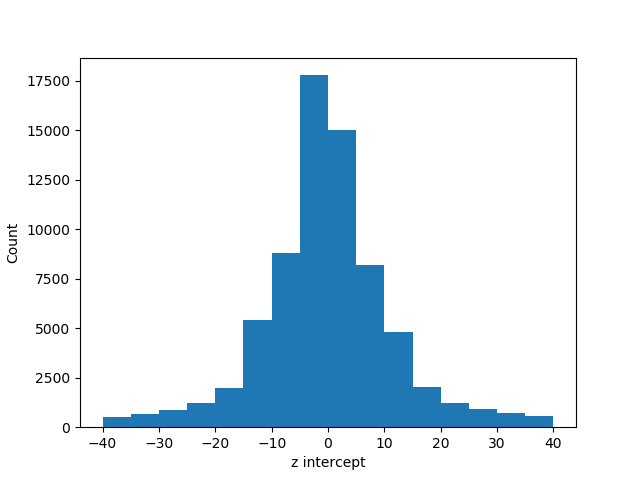}
\caption{Track segment $z$-intercepts (mm) in an event from the TrackML dataset. The standard deviation of the distribution is consistent with the size of the beamspot (5.5mm) within which tracks are produced.}
\label{fig:z_intercept}
\end{figure}

Furthermore, as particles are created in a small region (5.5~mm in the TrackML dataset~\cite{rousseau2018trackml}) along the $z$ axis close to the origin, segments are expected to point towards the origin in the $rz$-plane (figure~\ref{fig:z_intercept}). We model these expectations of the $z$-intercept by extrapolating pairs of line segments and applying a penalty if they do not intercept near the origin. To extrapolate the connected pair of track segments (for a more precise estimate of the $z$-intercept than extrapolating a single pair of hits), we consider the cross-term between $s_{ab}$ and $s_{ac}$ rather than a single track segment.
\begin{equation}
    \sum_{a,b,c} \left(z_c - \frac{z_c - z_a}{r_c - r_a} r_c\right)^\zeta s_{ab}s_{bc} .
\end{equation}

We propose a prior probability bias $P(s_{ab})$ of an edge being true based on its orientation in the $rz$-plane, adjusted by a constant inhibition term. Hence, we add a final term to our QUBO:
\begin{equation}
    \sum_{a,b}\left(\beta P(s_{ab}) - \gamma\right)s_{ab} ,
\end{equation}
where the prior probability $P(s_{ab})$ is calculated using a Gaussian kernel density estimation (KDE) of training data, described in more detail in section~\ref{sec:heuristics}.

The final QUBO incorporating all terms is given by:
\begin{align}
\label{equ:qubo}
\begin{split}
E = &-\sum_{a, b, c} \left(\frac{\cos^\lambda(\theta_{abc})+\rho \cos^\lambda(\phi_{abc})}{r_{ab} + r_{bc}}\right)s_{ab}s_{bc}\\
&+\eta\sum_{a, b c} \left(z_c - \frac{z_c - z_a}{r_c - r_a}r_c\right)^\zeta s_{ab}s_{bc}\\
&+\alpha\left(\sum_{b\neq c} s_{ab} s_{ac} + \sum_{a\neq c} s_{ab} s_{cb} \right)\\
&-\sum_{a,b}\left(\beta P(s_{ab}) - \gamma\right)s_{ab} .
\end{split}
\end{align}
The parameters are optimized by Bayesian optimization, sampling regions of the parameter space that are expected to provide the largest improvement in the objective function according to Bayesian inference. The optimization was run with 10 random starts to establish the initial prior probabilities, and then a total of 100 parameter sets were sampled on TrackML events with 500 particles/event to maximize the $F_1$ score (the harmonic mean between purity and efficiency) using SA. The optimal values are summarized in Table~\ref{tab:params}.
\newline
\begin{table}[H]
\centering
\caption{Parameters that enter the definition of the final QUBO (see equation~\ref{equ:qubo}).
The values are obtained using Bayesian optimization for best $F_1$ score optimizing the QUBO with SA.
The description corresponds to what term of the QUBO the parameters are driving.
}\label{tab:params}
\begin{tabular}{ c|c|c }
 Parameter & Value & Description\\ 
 \hline
 $\lambda$ & 13.17 & Track angle separator\\
 $\rho$ & 5.00 & High-momentum bias\\
 $\eta$ & 14.41 & Beam spot bias\\
 $\zeta$ & 1.79 & Beam spot separator\\
 $\alpha$ & 86.20 & Bifurcation penalty\\
 $\beta$ & 20.91 & Edge alignment penalty\\
 $\gamma$ & 9.79 & Total edge count penalty
\end{tabular}
\end{table}

\subsection{Datasets}
We use data from the TrackML Particle Tracking Challenge on Kaggle~\cite{rousseau2018trackml}, simulating the HL-LHC. The dataset consists of $8850$ events each consisting of approximately $10^5$ hits which cluster to about $10^4$ tracks. Around 15\% of the data is noise, with hits corresponding to no tracks. We use the spatial data along with the ground truth tracks to assess the performance and accuracy of the algorithm.

Single sensors are assembled with enough overlap to offer an hermetic coverage within each layer. Particle might therefore produce multiple hits per layer.
Duplicate hits can be removed empirically with geometrical considerations and minimal assumption on track parameters; we however use ground truth information for ease of processing. The additional hits can be added at limited extra cost during post-processing, so we consider this simplification of the dataset justified. Removing such closely-spaced hits effectively normalizes the distance between adjacent hits, allowing a single set of parameters to be chosen in the QUBO formulation.

Note that pattern recognition is performed in both the barrel and endcap detectors despite the higher density of tracks. Additionally, the detector geometry is more complex -- tracks no longer travel through layers sequentially in the barrel/endcap transition region -- requiring more complex heuristic methods described in section~\ref{sec:heuristics}.

\subsection{Heuristic pre-processing and problem decomposition methods}
\label{sec:heuristics}
To anneal an entire event at the HL-LHC, we would require a fully-connected quantum annealer with a qubit for each candidate edge. Given $10^5$ hits, this corresponds to a total of $10^{10}$ qubits (edges). This is well beyond the size of current and near-term quantum annealers, currently limited to a few thousand qubits. Similar issues are frequently encountered in other domains, and problem decomposition methods are therefore an important and active area of study in QA, based, e.g., on the belief propagation or divide-and-conquer algorithms~\cite{Bian:2016}. Here, to address the same need, we develop alternative heuristic methods with time complexity $O(h^2)$ where $h$ is the number of hits, to reduce the number of edges and hence the number of qubits needed. This ultimately allows events with $10^3$ to $10^4$ hits to be annealed on a quantum annealer with only 33 fully-connected logical qubits (see section~\ref{sec:scaling}). Since iterating over the data to construct a QUBO problem already runs in $O(h^2)$ time, these additional pre-processing heuristics do not significantly add computational time as the event size increases. Additionally, the complexity analysis is done without considering possible speed-up from parallel computation.

\begin{figure}[H]
\centering
\includegraphics[width = 0.4\textwidth]{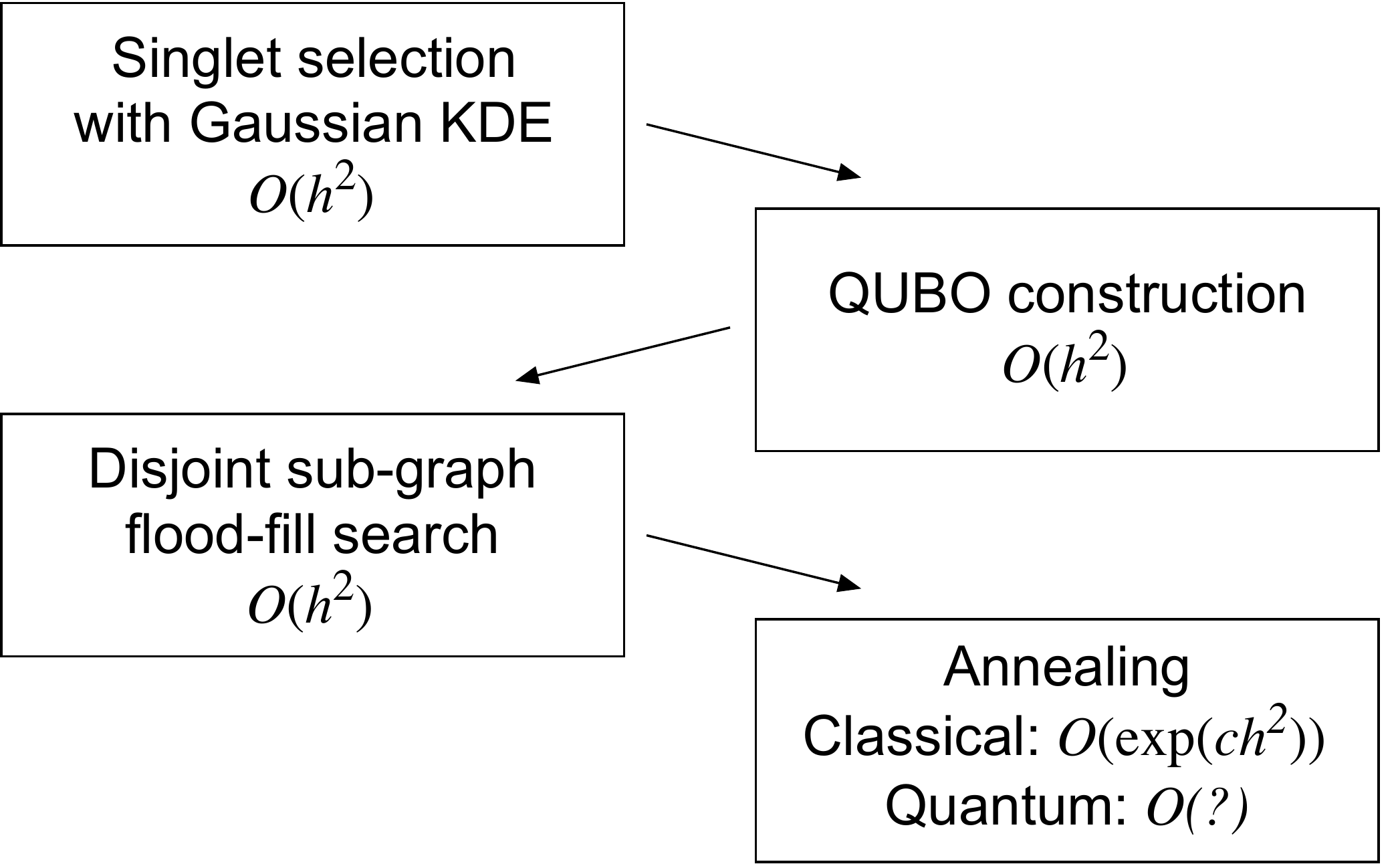}
\caption{Summary of the heuristic methodology for reconstruction. Each step in classical pre-processing has complexity $O\left(h^2\right)$ due to iterations over edges, where $h$ is the number of hits.}
\label{fig:pipeline}
\end{figure}

To limit the possible number of edges, we divide the event up into 32 overlapping sectors in the $xy$-plane, where each sector is $1/16^{\mathrm{th}}$ of the full azimuthal angle and half-overlaps with its neighboring sectors. In the TrackML dataset, we find that $>99\%$ of edges are within a single sector, and thus accurately solving individual sectors would guarantee correct reconstruction of over $99\%$ of the event in post-processing. 
We then apply the procedure consisting of selecting candidate edges with Gaussian kernel density estimation followed by subdividing the QUBO into smaller optimization problems (see figure~\ref{fig:pipeline}).

In the following we describe our two pre-processing (problem decomposition) methods.

\subsubsection{Gaussian kernel density estimation}
\begin{figure}[H]
\centering
\includegraphics[width=0.5\textwidth]{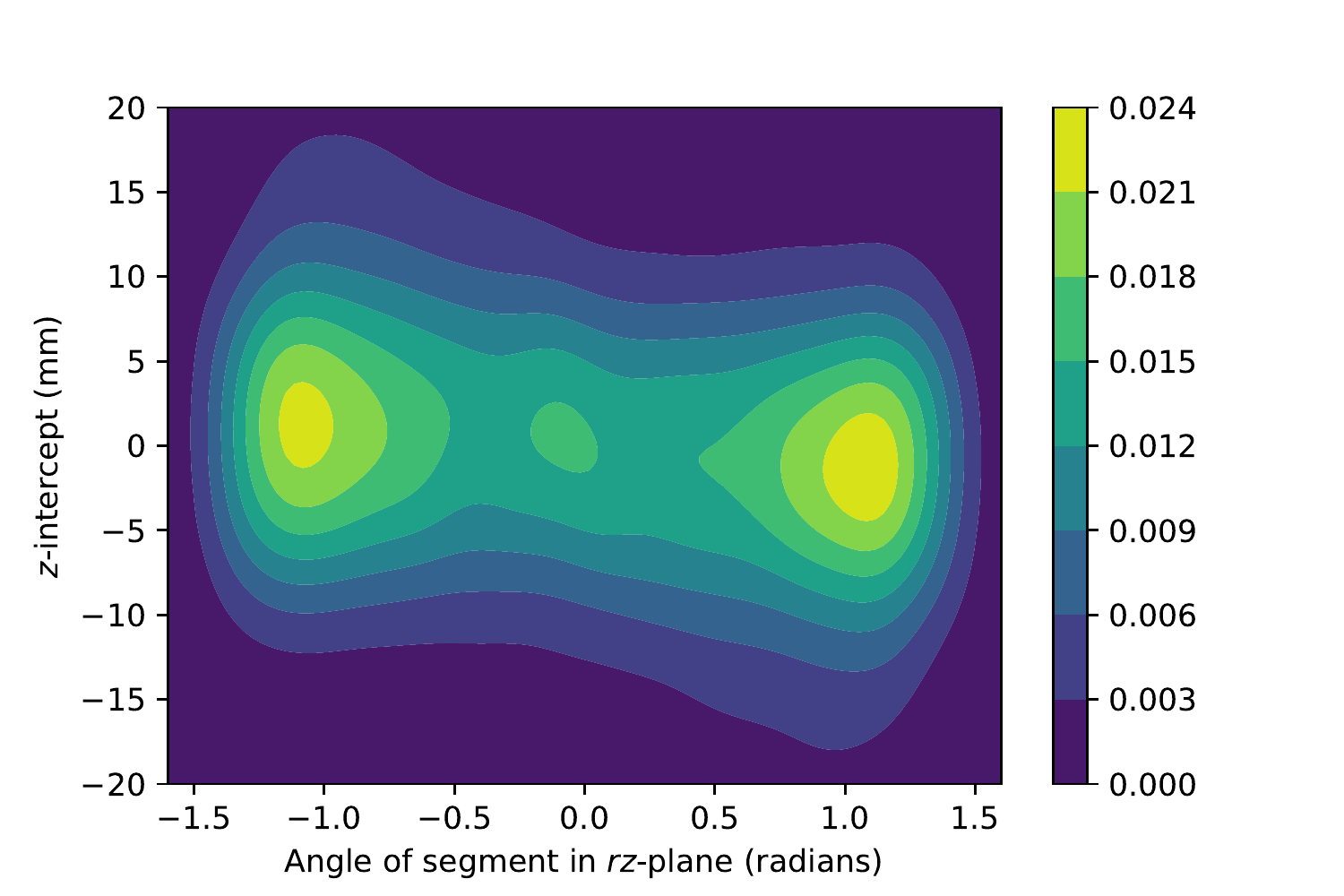}
\caption{Gaussian kernel density estimation of a prior probability for two hits to be connected by an edge, allowing information from the interaction points and detector geometry to be introduced into the QUBO.}
\label{fig:kde}
\end{figure}

To provide a general method for detector geometries beyond that of the TrackML dataset, we use Gaussian kernel density estimation (KDE) (figure~\ref{fig:kde}) to determine the prior probability that a given edge between two hits is true using data samples outside the test set. Since tracks typically originate from the interaction point near the origin, we train the Gaussian KDE on the $z$-intercept and the angle in the $rz$-plane of line segments based on ground truth in the TrackML data.

We apply a cut on the Gaussian KDE to reduce the size of the QUBO, yielding 93\% of all the true edges with approximately 1\% purity. Given $h$ hits, this has time complexity $O\left(h^2\right)$ as we traverse over all hits. We may then construct the QUBO outlined earlier, again traversing all edges with complexity $O\left(h^2\right)$.

\subsubsection{Sub-graphing}
Since we wish to anneal our problem using a small number of qubits, we further subdivide the problem into disjoint sub-graphs, separating individual communities of hits connected by edges. To do so, we perform a flood-fill search~\cite{torbert2016applied} to label each edge and prune the candidate edges from each node to only include the $5$ edges with the highest single-edge biases in the QUBO. Thus, this sub-division procedure also runs in time $O\left(h^2\right)$. We may then proceed to anneal the multiple QUBO problems with the number of problems scaling like the number of sub-graphs, i.e., as $O(h^2)$ since the sub-graphs divide the event into disjoint edge communities. The sub-graphing process is further detailed in section~\ref{sec:sparse} of the Supplementary Material.

\subsection{Annealing procedure}
\label{sec:scaling}
\begin{figure}[H]
\centering
\includegraphics[width=0.5\textwidth]{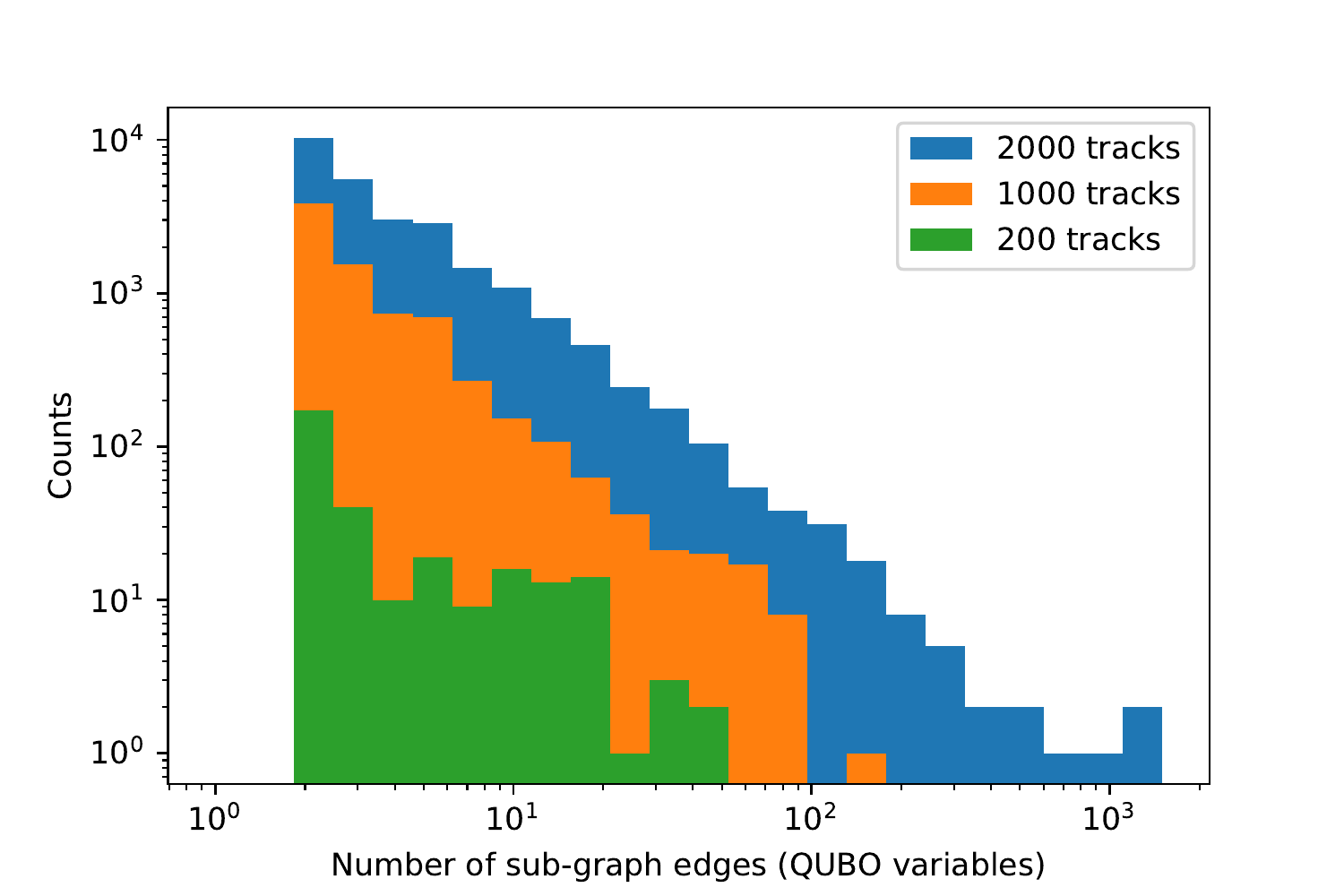}
\caption{Histogram of sub-graph sizes $m_i$ summed over the 5 largest event sectors (1/16 of an event) for different track densities. Each bin corresponds to the number of edges in a sub-graph, which is equivalent to the number of variables in a QUBO.}
\label{fig:histogram}
\end{figure}

Due to the QUBO construction of assigning each possible edge to a variable in the QUBO problem, we expect SA with no pre-processing to solve the tracking problem in exponential time with respect to the number of edges $h^2$, i.e., $O\left(\exp\left(ch^{2}\right)\right)$ 
for a constant $c>0$. 
After our sub-graphing procedure, we divide the event into $K = O(h^2)$ sub-graphs, and we expect total annealing time to grow as $\sum_{i=1}^{K} \exp\left(cm_i\right)$ where $m_i$ is the number of edges in sub-graph $i$. Hence, the overall scaling would depend on the distribution of $m_i$ (see figure~\ref{fig:histogram}).
However, since the sub-graphing procedure only reduces the complexity of the annealing (by dividing the larger QUBO into smaller sub-QUBOs), the procedure's complexity is bounded from above by $O\left(\exp\left(ch^{2}\right)\right)$. To verify this, we use SA and measure the convergence time as a function of the distribution of sub-graph sizes in section~\ref{sec:results2}. Details of the SA algorithm are provided in section~\ref{sec:sa} of the Supplementary Material. 

Although QA is not thought to generally yield a ground state solution to a QUBO problem in polynomial time, it may reduce the size of the constant $c$ in the time complexity $\sum_{i=1}^{K} \exp\left(cm_i\right)$, potentially offering a significant speedup over classical methods~\cite{qubotime,speedup}. To assess the possibility of a quantum speedup, we implement our procedure on a programmable quantum annealer built by D-Wave Systems Inc.~\cite{dwave} and housed at the University of Southern California's Information Sciences Institute. The D-Wave 2X architecture has $1,098$ superconducting flux qubits arranged in a Chimera graph, in which each qubit is coupled to at most $6$ others. To increase connectivity we perform a minor-embedding operation by mapping each QUBO problem onto ferromagnetic chains of qubits~\cite{embed1, embed2,klymko_adiabatic_2012,Cai:2014nx}; the result is a fully connected graph of 33 logical qubits, each of which is used to represent an edge.

We optimize the ratio between coupling within each chain to the largest coupling in the Hamiltonian to equal a factor of $3$. We find that this prevents chains from breaking (via noise from thermal excitations and domain walls) while still allowing qubits to flip to ensure that the transverse field Hamiltonian drives the dynamics~\cite{chains}. For each annealing run, we re-embed the problem $10$ times with randomized cross-term signs (gauges) to average out noise on local fields and couplers~\cite{job2018test}. For each gauge, we perform $10,000$ annealing runs before selecting the lowest-energy solution from all the outputs. Note that as the inherent noise in the annealing hardware improves in the future, fewer runs and gauges would be necessary. To test the effect of the annealing time (which in principle must be optimized in order to extract the true time to solution~\cite{PhysRevX.8.031016,speedup}), we compare runs from 5 to 800~$\mu$s.

\subsection{Benchmark studies}
To evaluate the performance of the annealing algorithm, we benchmark against random edge selection after pre-processing. Random edge selection simply randomly selects edges as true according to the expected fraction of true edge segments in the pre-processed data. Since the edge selection by annealing occurs after our heuristic edge selection with the Gaussian KDE and disjoint sub-graph search, comparison to random edge selection demonstrates that the patterns of hits are not found during pre-processing, but rather by solving the QUBO.

\section{Results}
\label{sec:results}
After measuring the overall tracking performance of our methodology, we present results on the scalability of our algorithm for both SA and QA to evaluate the possibility of a quantum speedup. We report error bars representing the 1 standard deviation ($\sigma$) spread of sector-by-sector purity and efficiency for TrackML events, indicating the robustness of the methodology.
Particle multiplicity and pileup are linearly dependent, where 2,000 particles per event corresponds to an average of 40 pileup.
\subsection{Tracking efficiency and purity}
\label{sec:results_qasa}
\begin{figure}[H]
\centering
\includegraphics[width=0.45\textwidth]{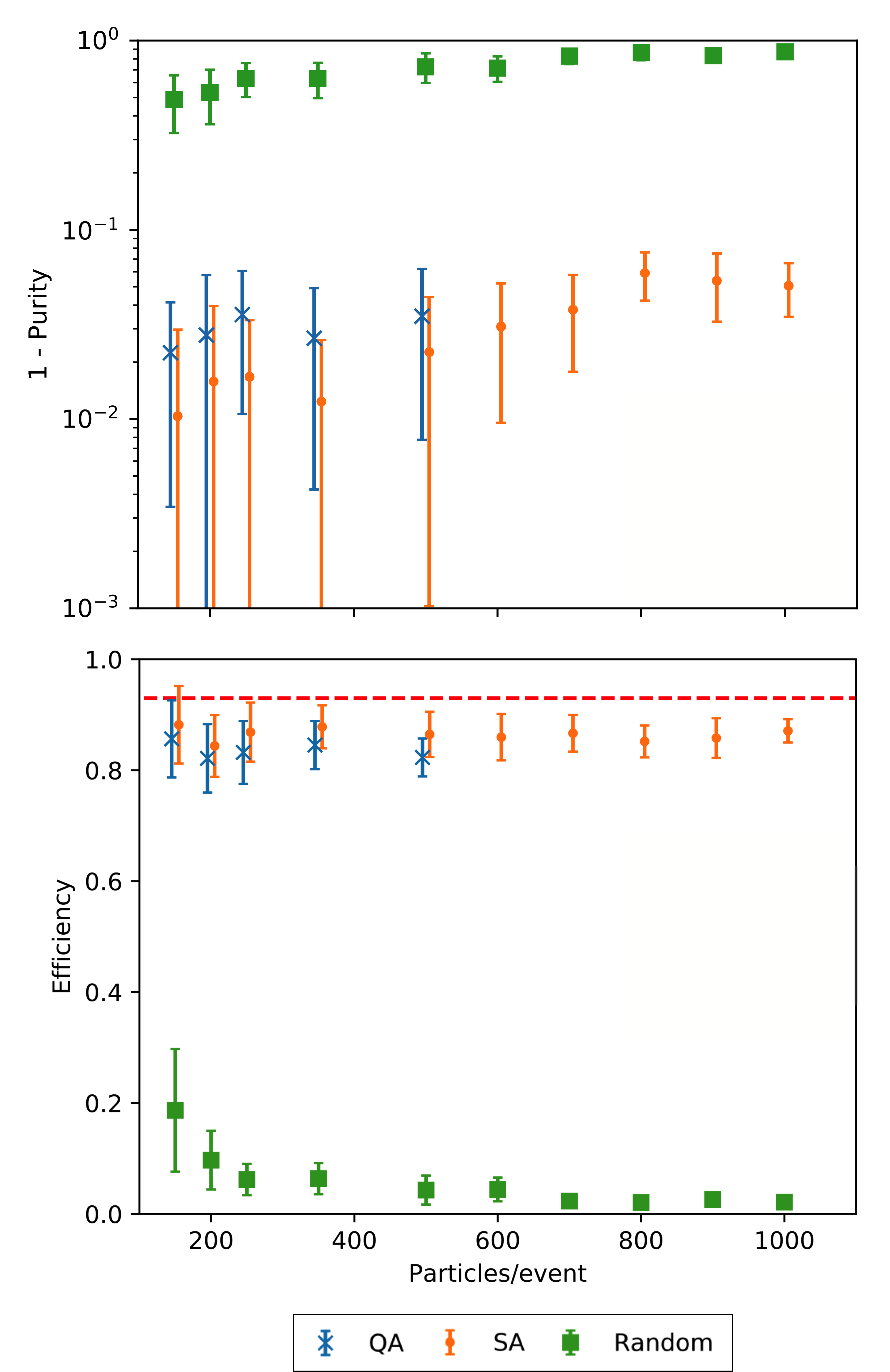}
\caption{QA and SA benchmarked against random annealing after pre-processing heuristics. All values are reported with $1\sigma$ error bars for tracks with at least 3 hits indicating the spread of event sectors. Additionally, the pre-processing places an upper bound of around 93\% efficiency (indicated by the dashed line).}
\label{fig10}
\end{figure}

\begin{figure}[H]
\centering
\includegraphics[width=0.5\textwidth]{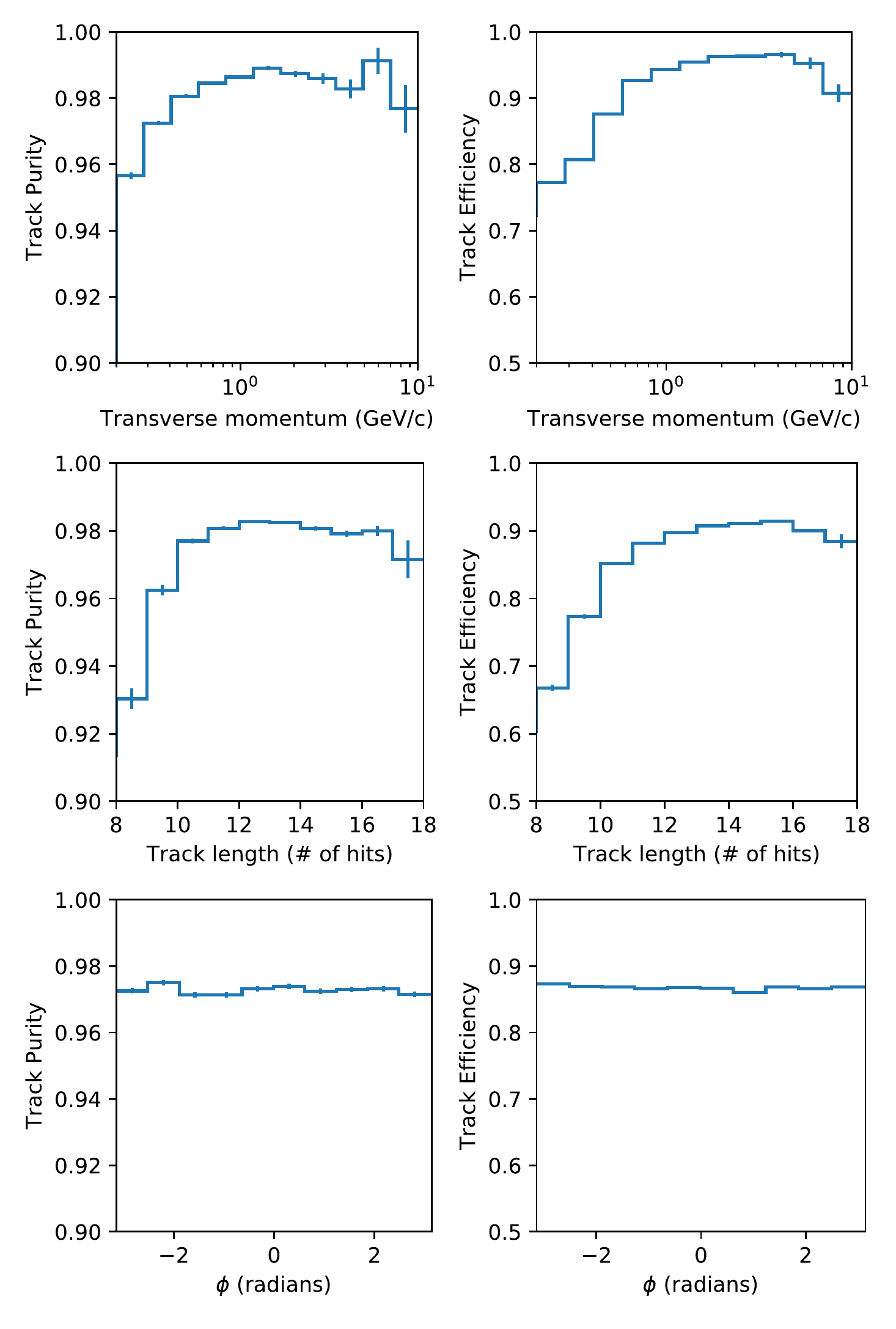}
\caption{Track purity and efficiency for SA results for events at 500 particles/event as a function of transverse momentum (top), track length (middle) and azimuthal angle (bottom).
}
\label{fig:pureff}
\end{figure}

To compare the QA and SA performance in terms of particle multiplicity (see figure~\ref{fig10}) and particle momentum (see figure~\ref{fig:pureff}), we use two metrics:
\begin{align*}
\text{Purity} &= \frac{\text{Number of true tracks reconstructed}}{\text{Number of tracks reconstructed}} ,\\
\text{Efficiency} &= \frac{\text{Number of true tracks reconstructed}}{\text{Number of true tracks}} .
\end{align*}
Due to the limited size of the D-Wave machine (33 fully connected logical qubits), we can only fit up to 500 tracks on the quantum annealer. However, to show that the performance of the algorithm does not significantly deteriorate at higher multiplicity, we include further results from SA.

As particle multiplicity increases, the random edge selection track efficiency and purity approach zero, while the SA and QA reconstructions maintain their performance. This suggests that the majority of tracking is completed in solving the QUBO rather than in our heuristic pre-processing methods. Although quantum annealing on D-Wave hardware does not outperform SA, it consistently obtains a solution of similar quality. The SA algorithm's slightly better performance may be attributable to a lack of noise in embedding the Hamiltonian as well as the ability to fully encode the problem without chains of qubits that cause additional error in the readout process.

\begin{figure}[H]
\centering
\includegraphics[width=0.4\textwidth]{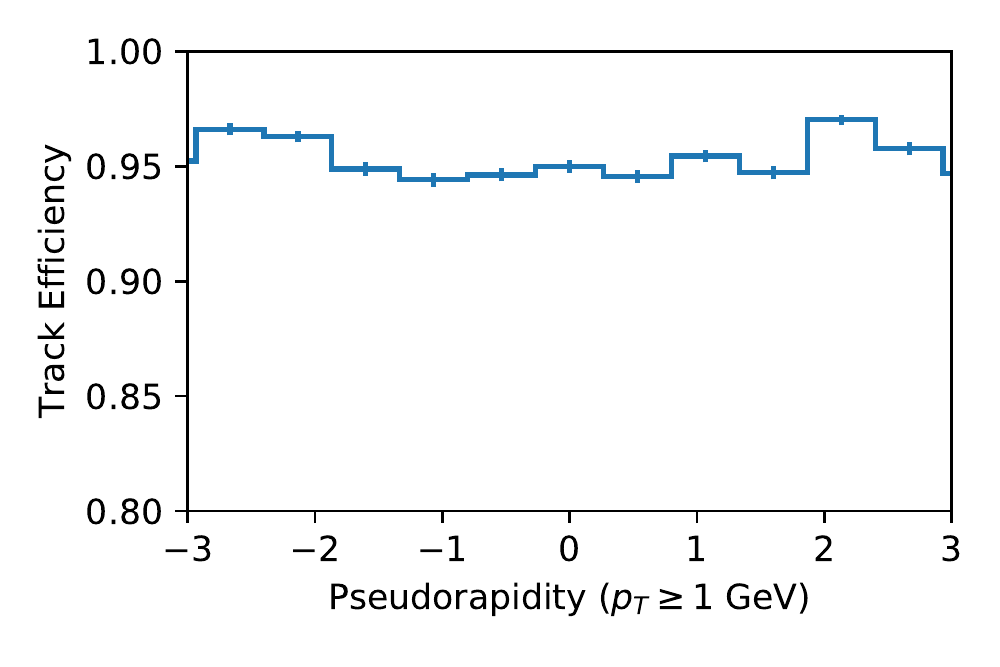}
\caption{Efficiency as a function of $\eta$ for tracks with $p_\mathrm{T} >$ 1~GeV. The track distribution is typically constant in $\eta$.}
\label{fig:ps}
\end{figure}

We present the performance in terms of track efficiency and purity across several physical variables (see figures~\ref{fig:pureff} and~\ref{fig:ps}). 
For reference, 96\% of true edge segments in the TrackML dataset belong to tracks that are 8 to 18 hits in length. Metrics are calculated for tracks at least 3 hits in length. Only SA was used in the figures to improve the statistical uncertainty with a larger number of events.

Track reconstruction performance increases with transverse momentum $p_\mathrm{T}$, recording higher-momentum particles with both higher efficiency and purity since tracks are straighter and thus better-suited to the QUBO formulation. 
The drop in efficiency at high $p_\mathrm{T}$ is observed in many solutions of the TrackML challenge~\cite{rousseau2018trackml} and might be an artifact of the dataset simulation.
Moreover, tracking performance remains constant with azimuthal angle $\phi$ in the $xy$-plane, indicating the consistency of the tracking algorithm since events are typically homogeneous in $\phi$. Similarly, we find that the tracking algorithm shows consistent performance across the full range of pseudorapidity $\eta = -\log \tan \frac{\theta}{2}$, where $\theta$ is the polar angle between particle momentum and the beam axis (see figure~\ref{fig:ps}).

\subsection{Impact of pre-processing on SA performance}
\label{sec:results2}

\begin{figure}[H]
\centering
\includegraphics[width=0.5\textwidth]{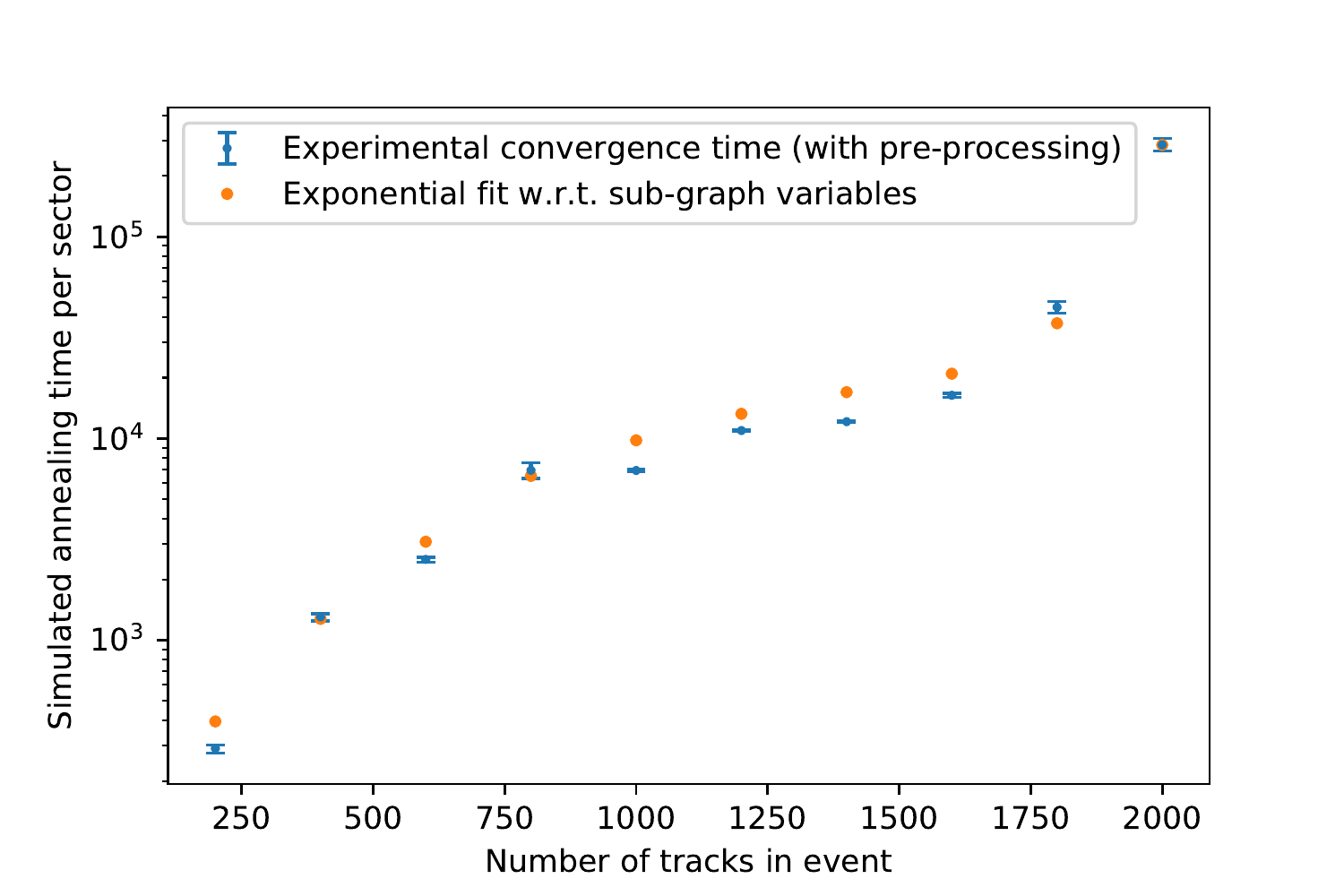}
\caption{SA convergence time with a fit $O\left(\sum_{i}^{K} \exp\left(cm_i\right)\right)$. By assuming the logarithm of convergence time is proportional to the distribution of sub-graph sizes $\{m_i\}$, we find a reasonable model for pre-processed scaling with a proportionality constant of $c = \left(8.52 \pm 0.076\right)\times 10^{-3}$. Error bars show the $1\sigma$ variation in annealing time, not including variation across events.}
\label{fig:sa_scale}
\end{figure}

With SA, we expect the time scaling after pre-processing to increase exponentially as $O\left(\sum_{i=1}^{K} \exp\left(cm_i\right)\right)$ for $c>0$ where the $i^{\mathrm{th}}$ sub-graph has $m_i$ edges. 
This exponential-time fit with respect to the number of sub-graph edges (or equivalently, the number of QUBO variables) appropriately models the computational time of SA with $c = \left(8.52 \pm 0.076\right)\times 10^{-3}$, as seen in figure~\ref{fig:sa_scale} where the quality of the fit is shown as a function of track number.

\begin{figure}[H]
\centering
\includegraphics[width=0.5\textwidth]{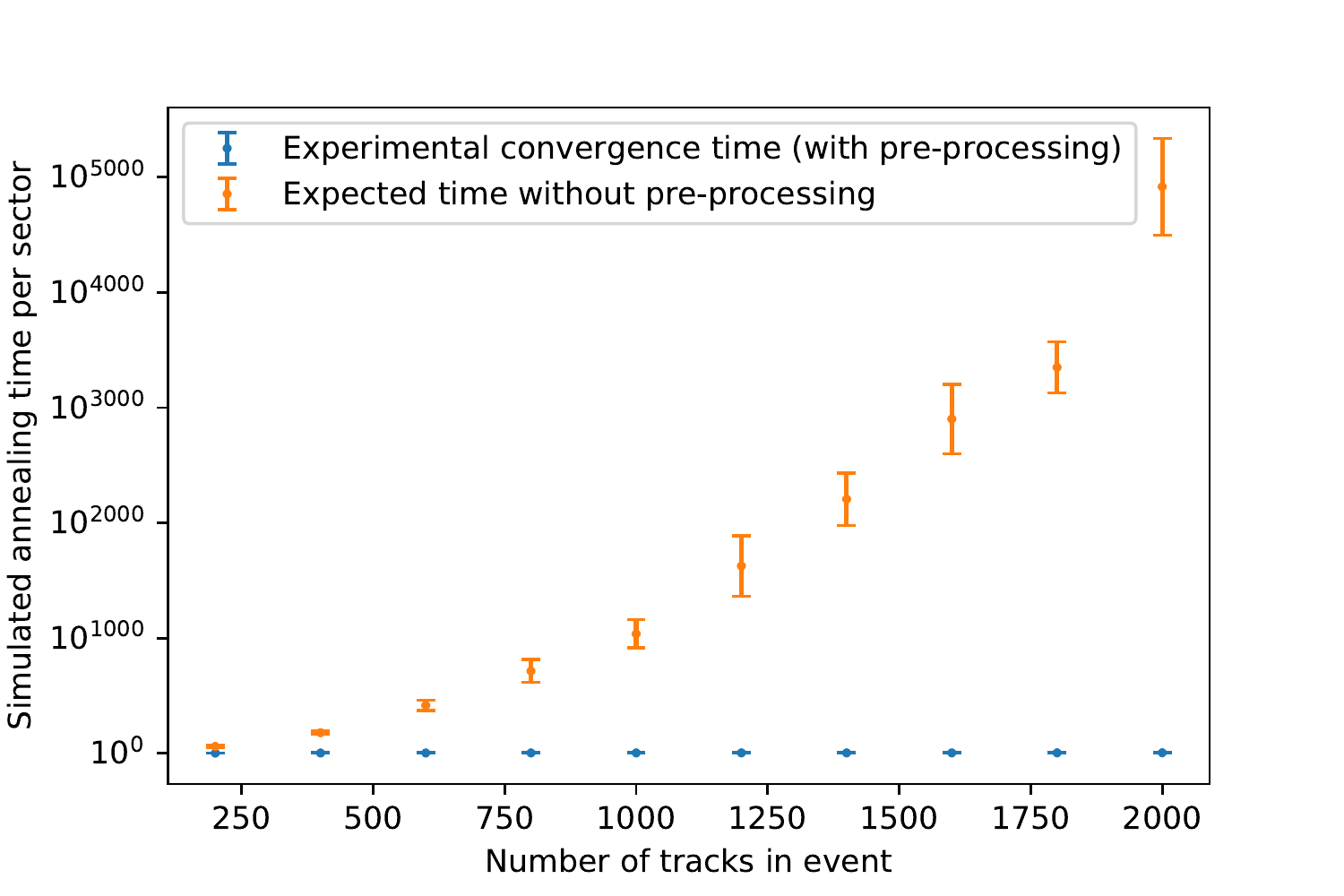}
\caption{SA convergence time bounded from above by $O\left(\exp\left(ch^{2}\right)\right)$, with $c = \left(8.52 \pm 0.076\right)\times 10^{-3}$.}
\label{fig:bounded}
\end{figure}

Note that although the convergence time is exponential after pre-processing, this is bounded from above by $O\left(\exp\left(ch^{2}\right)\right)$ obtained if there were no pre-processing (see figure~\ref{fig:bounded}).

\subsection{Feasibility of quantum speedup}
\label{sec:QAvsta}

In general, it is unlikely that QA can achieve polynomial time on this problem, but
there is room for a potential quantum speedup if QA can reduce the exponent $c$.

\begin{figure}[H]
\centering
\includegraphics[width=0.5\textwidth]{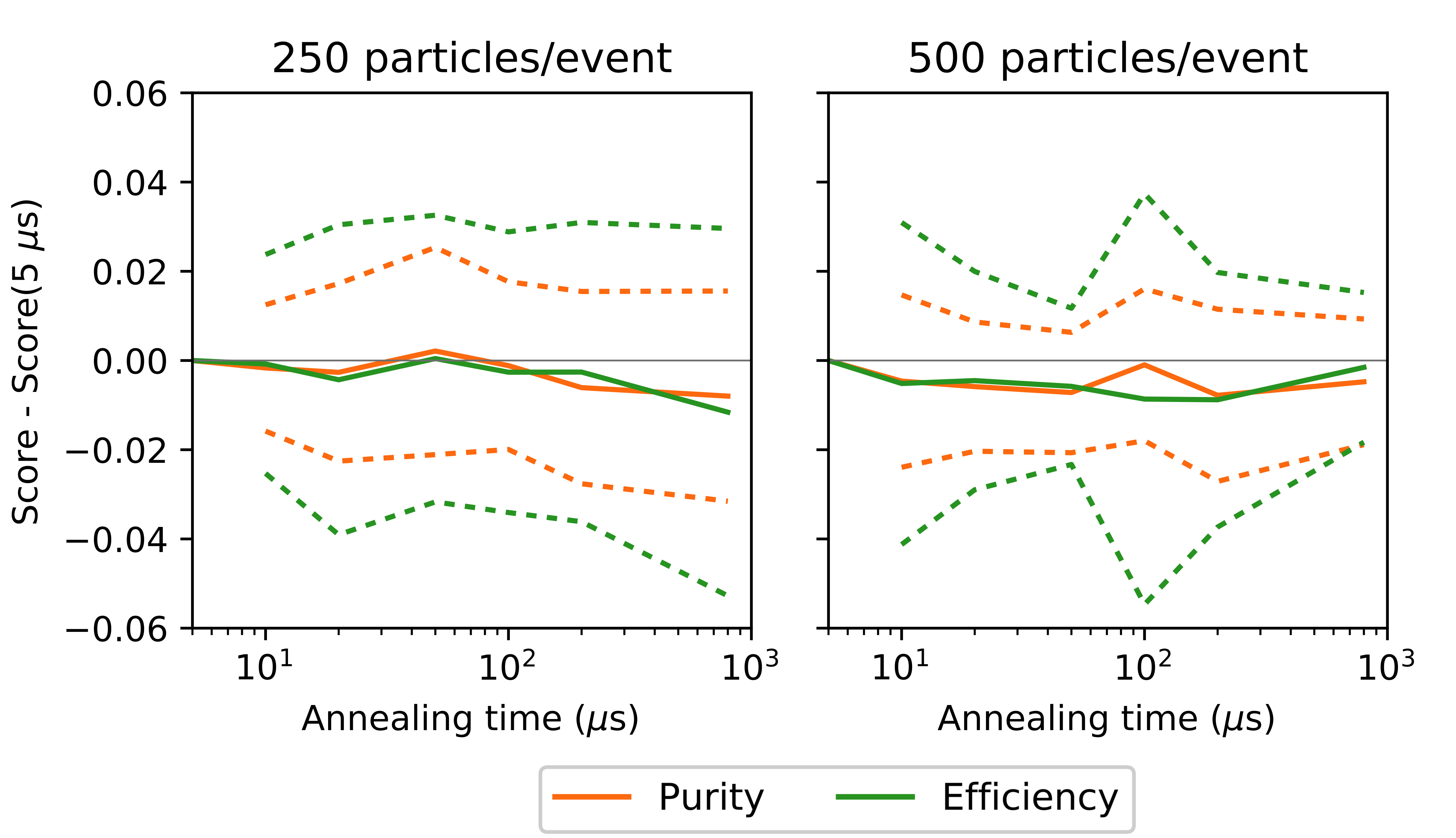}
\caption{Comparison of the purity and efficiency for different annealing times (5 to 800~$\mu$s) on the D-Wave 2X with respect to the purity and efficiency for a 5~$\mu$s anneal time. The dashed lines show $1\sigma$ error bars calculated from the sector-by-sector distribution of differences between the purity or efficiency for a given anneal time and the 5 $\mu$s anneal time.
}
\label{fig:diff}
\end{figure}

To fully test a quantum speedup~\cite{speedup}, a complete analysis of the problem scaling would require identifying the optimal anneal time for each problem size~\cite{PhysRevX.8.031016}, which we did not attempt other than a crude sampling of several annealing times (see figure~\ref{fig:diff}).
Quantum annealing yields very similar results with no clear trend using various annealing times (5, 20, 50, 100, 200 and 800~$\mu$s), suggesting that, unfortunately, we lack data at sufficiently small timescales to fully determine the time scaling while using D-Wave since the QUBO problems are satisfactorily solved by the shortest possible anneal time allowed by the hardware (5~$\mu$s) even for the largest events.
Indeed, contrary to SA, the performance on D-Wave deteriorates slightly with additional annealing time, possibly due to the effects of $1/f$ noise, in which low frequency components impact performance at long anneal times~\cite{DW-white-paper-1overf}. 

To fully assess the quantum speedup question would require a larger programmable quantum annealer such that a larger QUBO can be encoded, resulting in a 
minimum as a function of problem size in the time to solution when annealing. Given the scaling of the QUBO size as $O(h^2)$ and our limit of 500 tracks on the 33 fully-connected logical qubits of the D-Wave 2X, an event with 10,000 tracks (corresponding to HL-LHC conditions) would have a factor of 400 more edges and thus we expect that a programmable quantum annealer with a similar architecture to D-Wave must have approximately 10,000 fully-connected qubits to fully process a sub-graphed event at the HL-LHC. This is twice as many qubits as in the next generation D-Wave processor based on the (increased connectivity but not fully-connected) Pegasus architecture~\cite{DWave-Pegasus-techreport}.

\section{Related work}
\label{sec:related}
There is limited research on performing track reconstruction using quantum devices or quantum algorithms. The work reported in this document is one of the first of its kind.

Track reconstruction using quantum annealing has been explored in Ref.~\cite{bapst2019pattern}, using triplets of hits (as opposed to doublets in our approach) which increases pre-processing time to $O(h^3)$.
However, the solution proposed by the authors is limited to tracking in a simpler detector sub-region of simulated LHC data. It uses extensive classical pre-processing using the ATLAS track-seeding code and internal QUBO-solving methods in the D-Wave API that use further classical pre-processing and have higher overhead. In contrast, we require no track seeding and demonstrate tracking in both the barrel and endcap regions, explicitly controlling the annealing methodology in solving the QUBO. This allows us to show that the annealing procedure is computationally responsible for the majority of particle tracking, not the pre-processing methods.

The application of quantum associative memory for track pattern recognition and its circuit-based implementation on current hardware has previously been theoretically explored by members of the same team~\cite{shapoval2019quantum}.
Unlike our approach, the quantum associative memory framework is completely supervised: track candidates are tested by comparing them to simulated track patterns, which must be stored in quantum memory. Because we use a parameterized QUBO formulation of the tracking problem, we do not need to store simulated track data. On the contrary, our approach is based on physical models where the weights of the QUBO are guided by both physical expectations and simulated data.

Finally, quantum annealing has been proposed for a different but closely related problem of vertex reconstruction~\cite{das2019track}. 
However, this application is limited to events with up to 15 particle tracks, and it does not aim at reconstructing tracks, but rather aggregates them in a fixed number of vertices. 

\section{Discussion}
\label{sec:conclusion}
We find that charged particle tracking can be successfully interpreted as a segment classification problem in a quadratic unconstrained binary optimization (QUBO) framework, using efficient classical pre-processing followed by quantum or simulated annealing. By using features such as track helicity, momentum, interaction point position, and edge alignment bias as predicted by Gaussian kernel density estimation, the Denby-Peterson framework~\cite{denby1988neural, peterson1989track, stimpfl1991fast} may be adapted to LHC conditions. Although current annealing hardware limitations impose stringent constraints on the size of the optimization problem, we propose a methodology to systematically reduce the size of the QUBO by finding disjoint sub-graphs and performing multiple iterations of annealing. Ultimately, our work indicates that tracking problems at the High-Luminosity LHC may be studied with competitive efficiency and purity results on programmable quantum annealers in the future, while the question of a quantum speedup in this context remains open.

Besides providing potential applications for quantum annealing, our methodology establishes the utility of classical simulated annealing for modern tracking problems, and may thus be run on high-performance Field Programmable Gate Array (FPGA) simulated annealing hardware~\cite{fujitsu} with up to 8192 bits, as well as the Coherent Ising Machine~\cite{Inagaki:2016aa} with 2000 fully connected spins.
As these are classical annealing approaches they are of course not expected to exhibit a quantum speedup. However, they are fully connected, overcoming embedding challenges associated with the D-Wave annealer and enabling larger problems to be encoded with fewer bits. By exploiting the performance advantages of FPGAs and classical annealers, one could perform preliminary tracking at the trigger level. Furthermore, instead of tuning the QUBO parameters to maximize the harmonic mean of track efficiency and purity, the QUBO parameters may be tuned for either high track efficiency (to reduce overall data size) or high track purity (to eliminate entire tracks from the dataset) before applying traditional tracking methods (such as Kalman filters).
The approach presented in this paper could be used as a first step of an iterative tracking procedure, otherwise already in use in experiments like CMS.

We note that the spin states found by the D-Wave annealer suggest that sufficiently good solutions to the tracking problem QUBO may be found by programmable quantum annealers without fully solving the QUBO for its ground state. Thus, despite not directly identifying a quantum speedup in this work, we conclude that there remains practical potential of quantum annealing for charged particle tracking. Moreover, as of the time of writing, quantum annealing is the only quantum hardware approach that can accommodate tracking problems large enough to be of any practical interest.

\section*{Acknowledgments}
Part of this work was conducted at  ``\textit{iBanks},'' the AI GPU cluster at Caltech. We acknowledge NVIDIA, SuperMicro and the Kavli Foundation for their support of ``\textit{iBanks}.''
This work is partially supported by DOE/HEP QuantISED program grant, Quantum Machine Learning and Quantum Computation Frameworks (QMLQCF) for HEP, award number DE-SC0019227.
JMD is supported by Fermi Research Alliance, LLC under Contract No. DE-AC02-07CH11359 with the U.S. Department of Energy, Office of Science, Office of High Energy Physics.  The work is also supported in part by the AT\&T Foundry Innovation Centers through INQNET, a program for accelerating quantum technologies. The work of DL and JJ was partially supported by the Office of the Director of National Intelligence (ODNI), Intelligence Advanced Research Projects Activity (IARPA), via the U.S. Army Research Office contract W911NF-17-C-0050.

\printbibliography
\end{multicols}
\newpage

\section*{Supplementary Material}
\renewcommand{\thesubsection}{\arabic{subsection}}
\subsection{Energy analysis}
To establish the quality of the QUBO construction, we evaluate the Ising model energy
\begin{align}
\begin{split}
E = &-\sum_{a, b, c} \left(\frac{\cos^\lambda(\theta_{abc})+\rho \cos^\lambda(\phi_{abc})}{r_{ab} + r_{bc}}\right)s_{ab}s_{bc} +\eta\sum_{a, b c} \left(z_c - \frac{z_c - z_a}{r_c - r_a}r_c\right)^\zeta s_{ab}s_{bc}\\
&+\alpha\left(\sum_{b\neq c} s_{ab} s_{ac} + \sum_{a\neq c} s_{ab} s_{cb} \right) +\sum_{a,b}\left(\beta P(s_{ab}) - \gamma\right)s_{ab}
\end{split}
\label{sup-qubo}
\end{align}
for a 500-track sector and compare the minimum-energy solution to the ground truth. Since the QUBO formulation does not allow the track efficiency and purity to be directly optimized, the QUBO construction provides an indirect objective function that should be highly correlated with precision and recall.
\begin{figure}[H]
\centering
\includegraphics[width=0.8\textwidth]{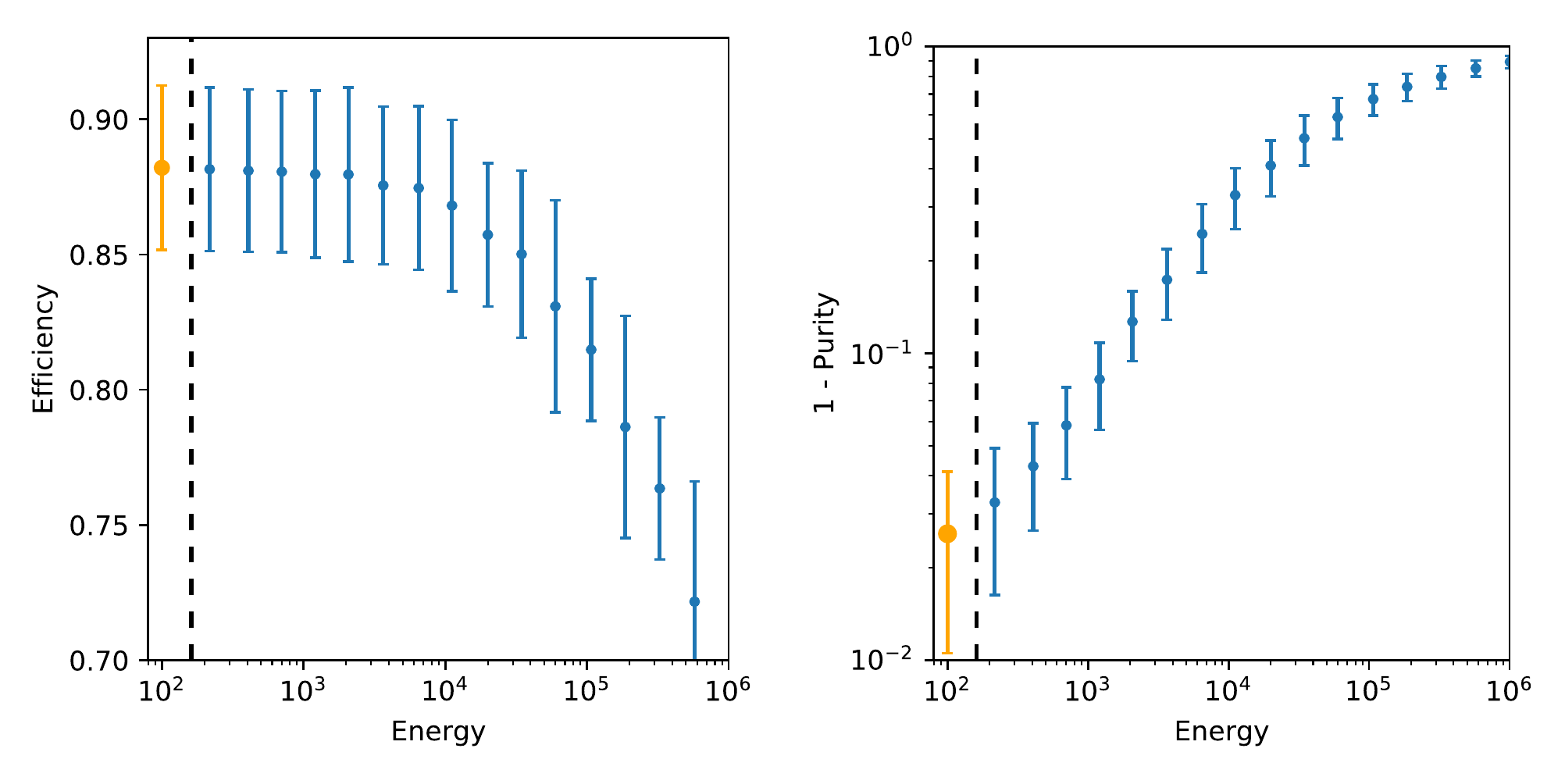}
\caption{Correlation of Ising model energy with track efficiency and purity for a 500-track sector. The Hamiltonian has been offset to normalize the lowest energy (orange) to 1000. Ground truth energy (100\% purity and 93\% efficiency) is shown by the black dashed line.}
\label{fig:energy}
\end{figure}
As seen in figure~\ref{fig:energy}, the ground truth energy is in the neighborhood of the minimum energy relative to the purity and efficiency metrics, providing further confirmation that the QUBO construction accurately represents the tracking problem.

\subsection{Sparse QUBO construction}
\label{sec:sparse}
\begin{figure}[H]
\centering
\includegraphics[width=0.8\textwidth]{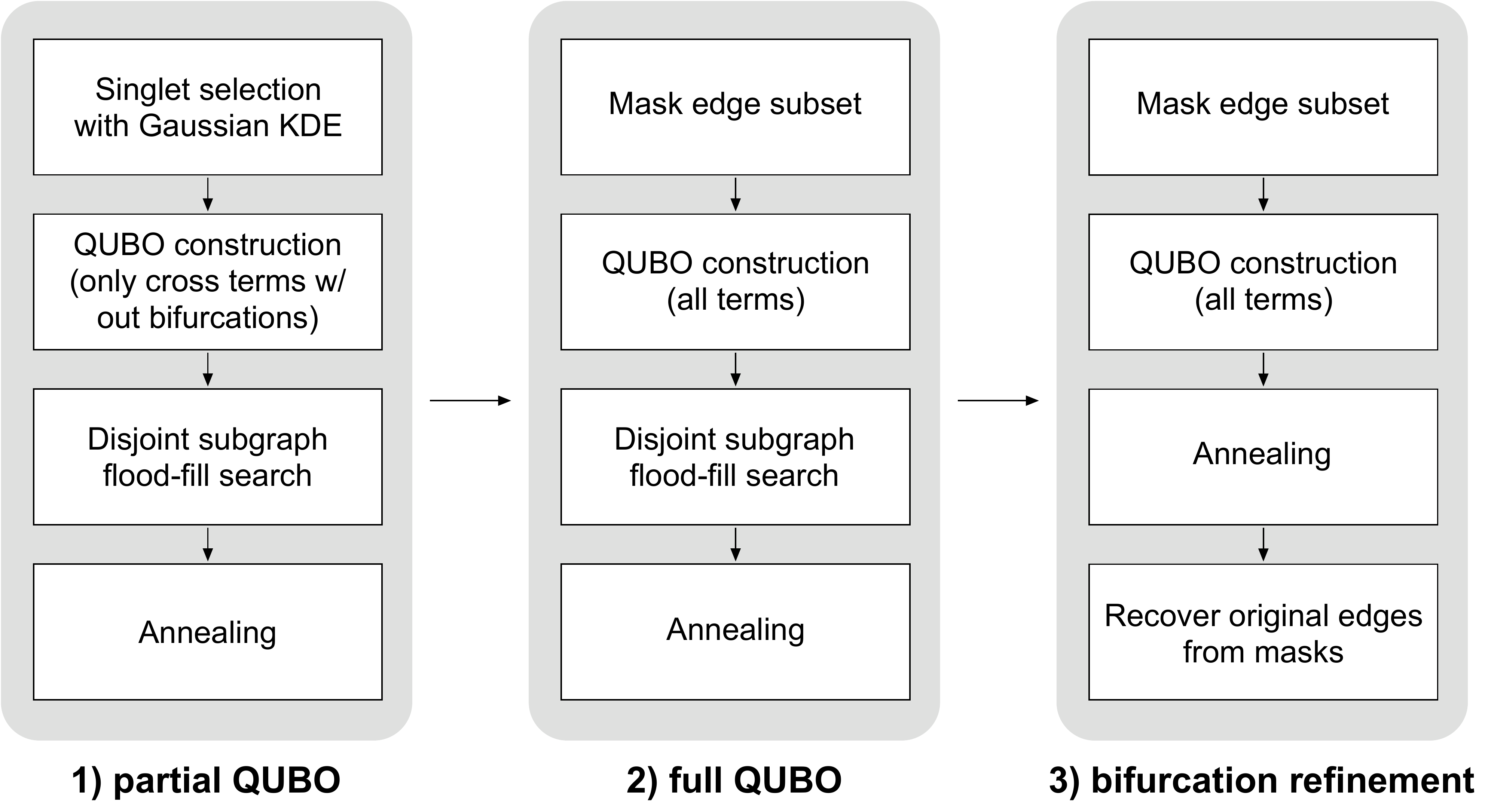}
\caption{Summary of dimension reduction methodology. After solving a partial QUBO with no bifurcation or single-spin terms, quantum annealing may be performed on the full QUBO using a partially connected and then fully connected graph. By tuning the cutoff parameters at each stage (i.e. constraining the sizes of edge subsets), the problem can be scaled to quantum hardware while attempting to preserve the quality of the solution.}
\label{fig:dwave-pipeline}
\end{figure}

\begin{figure}[H]
\centering
\includegraphics[width=0.65\textwidth]{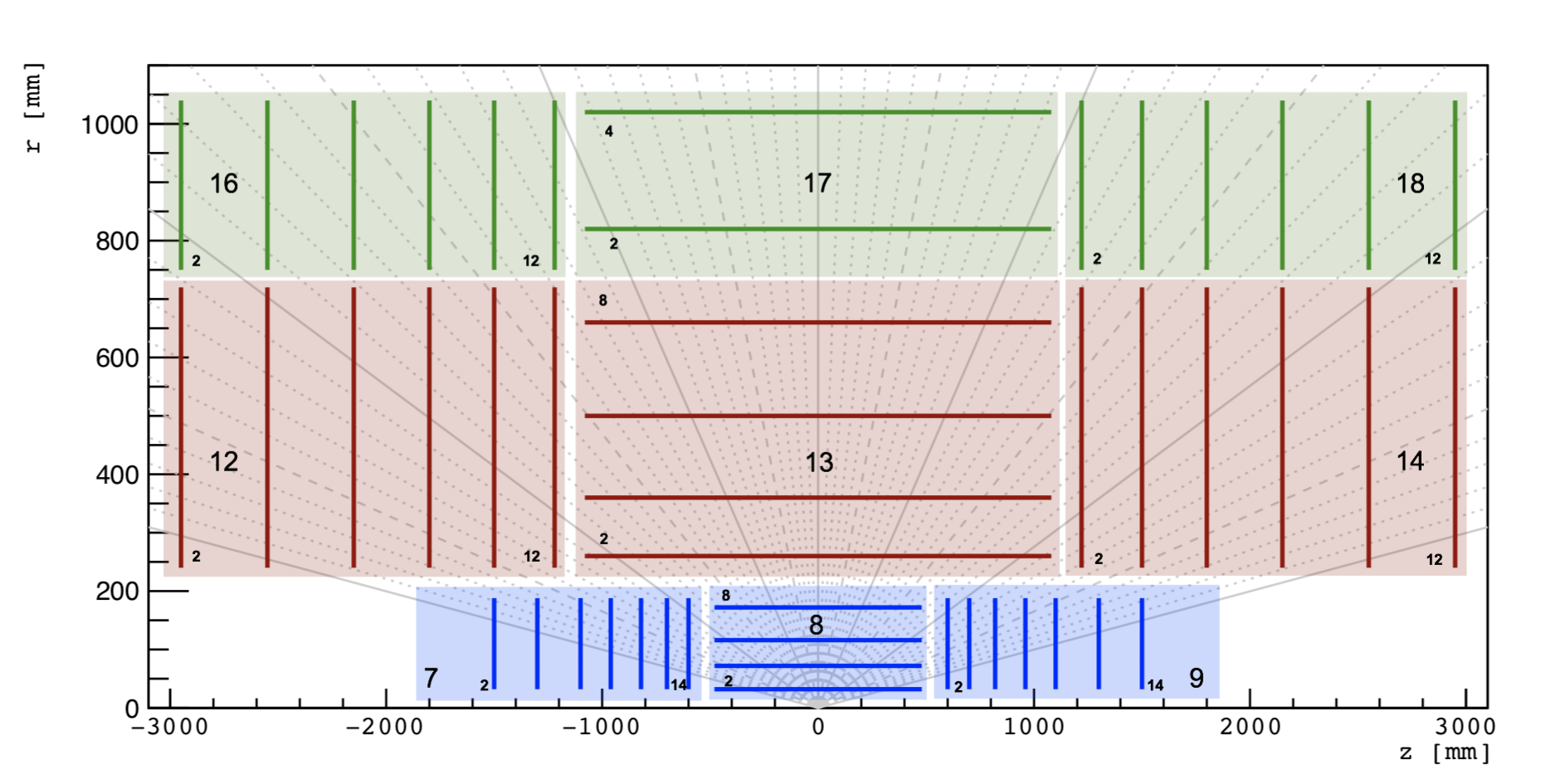}
\caption{TrackML detector geometry in the $rz$-plane \cite{rousseau2018trackml}.}
\label{fig:detector}
\end{figure}

Due to the lack of full connectivity on current programmable quantum annealers, we propose additional methodology (see figure~\ref{fig:dwave-pipeline}) to increase sparsity in the quadratic unconstrained binary optimization (QUBO) formulation of the tracking problem.
In the first stage, a Gaussian kernel density estimator (KDE) is used to approximate a prior probability on a given edge being on or off based on the angle in the $rz$-plane and $z$-intercept of a line segment between two hits (a ``singlet''). Although geometric approaches may be taken in place of a Gaussian KDE, the complex geometry of the detector end-caps (right and left regions of figure \ref{fig:detector}) as well as the versatility in extending the methodology to new detector geometries provide justification for using a generic prior probability estimation methodology.
By placing a threshold on the minimum prior probability, the candidate set of edges can be classically reduced in size. A QUBO is then constructed from edge compatibility scores without bifurcations or bias terms:
\begin{align}
\begin{split}
E = &-\sum_{a, b, c} \left(\frac{\cos^\lambda(\theta_{abc})+\rho \cos^\lambda(\phi_{abc})}{r_{ab} + r_{bc}}\right)s_{ab}s_{bc} + \eta\sum_{a, b c} \left(z_c - \frac{z_c - z_a}{r_c - r_a}r_c\right)^\zeta s_{ab}s_{bc}
\end{split}
\end{align}
Since the remaining terms in the QUBO are negative, solving this QUBO yields high-efficiency but low-purity tracks. To reduce connectivity, we perform a disjoint sub-graph flood-fill search on the space of edges, labeling up to the 5 best edge connections of a starting edge and spreading the label of each of the children edges to their neighbors. Thus, each edge is connected to at most 5 other edges, allowing the problem to be more easily embedded in the Chimera graph architecture of the D-Wave annealer since each edge is encoded as a qubit. Moreover, the event is divided into sub-graphs that are disjoint by edges, allowing it to be annealed in multiple QUBO problems.

Since the first stage retains efficiency, we may use the subset of edges it requires to be on in the second stage while discarding all edges classified as off. The full QUBO (Eq.~\ref{sup-qubo}) is then annealed on a sub-graphed event, following the same procedure as outlined above. Although this further increases purity, the sub-graphing divides the event by edges, preventing all bifurcations from being evaluated.

Hence, a final annealing on the new edge subset is applied to the full QUBO without any sub-graph constraints. This ensures that all possible bifurcations are included in the annealing process, yielding a QUBO that approximates the full event by only considering high-quality edges determined by the preceding annealings.

\begin{figure}[H]
\centering
\includegraphics[width=0.4\textwidth]{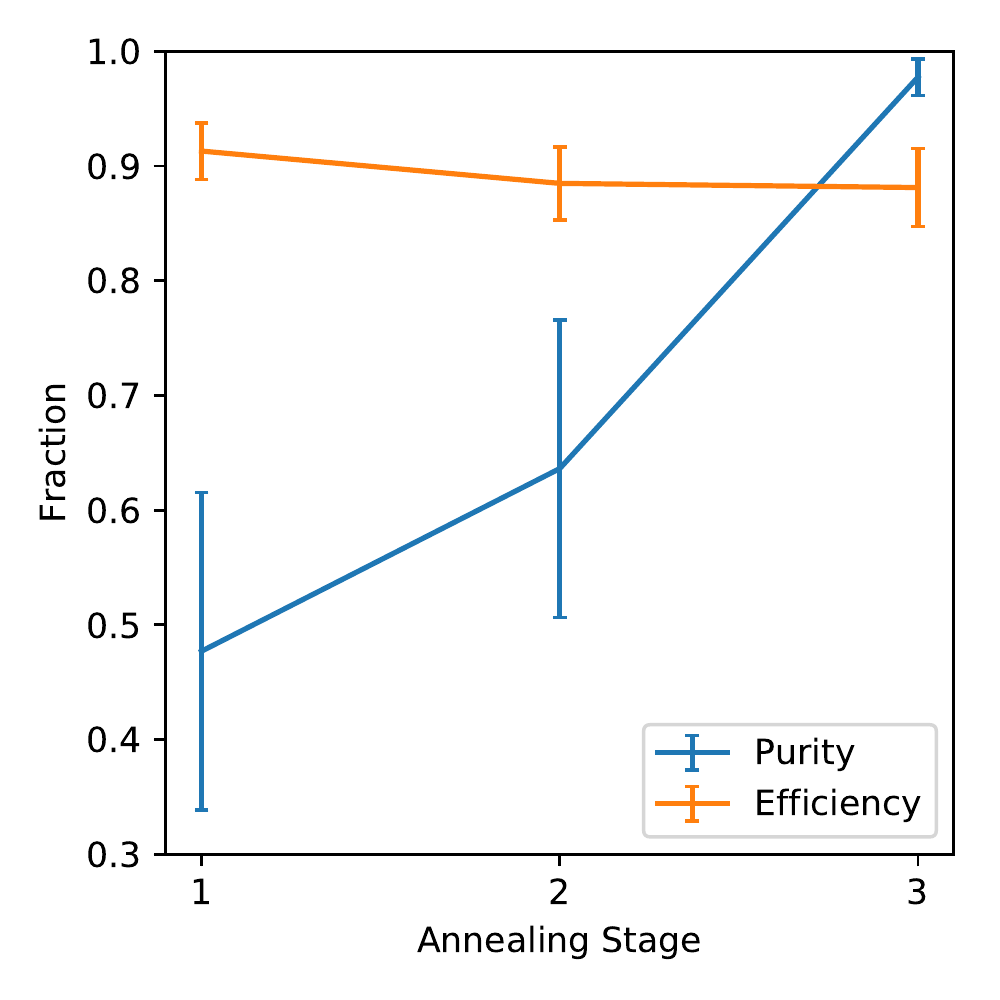}
\caption{Purity and efficiency dependence on annealing stage in the sparse QUBO construction. Purity is significantly improved without a large impact on overall tracking efficiency (maximum 93\% from the Gaussian KDE cutoff).}
\label{fig:iterations}
\end{figure}

\begin{figure}[H]
\centering
\includegraphics[width=0.95\textwidth]{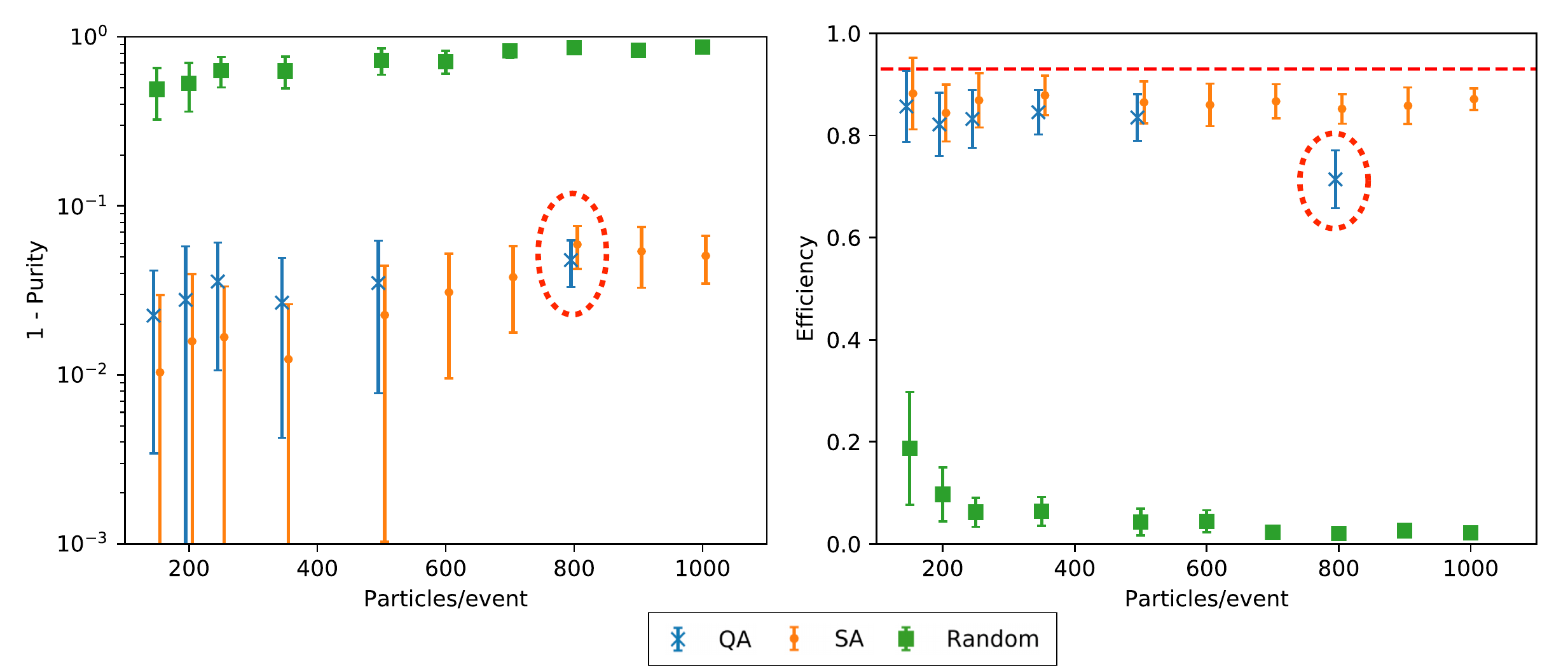}
\caption{Purity and efficiency for with QUBO construction parameters tuned to process larger events with high purity but lower efficiency. The QA point for 800 tracks (circled) is shown with the new parameters, while the remaining data uses the parameters reported in the main text.}
\label{fig:moretracks}
\end{figure}

For the parameters reported in the main text, this pre-processing procedure with sub-graphs enables events of up to 500 tracks to be annealed on the D-Wave 2X, corresponding to encoding $60,000$ QUBO variables in a fully-connected graph on an annealer with only 33 fully-connected logical qubits. By changing the parameters of the QUBO construction, larger events may be annealed with further cost to performance. For instance, we may modify the threshold of the edge affinity term, which sets the reward weight to 0 if $\cos^\lambda(\theta_{abc}) < \tau$, i.e. if two adjacent edge segments do not sufficiently line up to a helix (shown in figure \ref{fig3}). As $\tau$ increases, the cross-terms are increasingly sparse, allowing the event to be broken into a larger number of sub-graphs. By increasing $\tau=0.996$ (used in the main text) to $\tau=0.9997$, we may fit events as large as 800 tracks on the D-Wave 2X, corresponding to encoding $160,000$ QUBO variables in a fully-connected graph. However, this decreases overall performance (see figure~\ref{fig:moretracks}).

\subsection{Simulated annealing}
\label{sec:sa}

\newcommand{\pvec}[1]{\vec{#1}\mkern2mu\vphantom{#1}}

In simulated annealing (SA)~\cite{kirkpatrick1983optimization}, we start with a randomly initialized vector of spins $\vec{s}$ of length $N$. In each sweep, we flip $N$ variables one by one. If the flip decreases the QUBO energy, $E(\pvec{s}') < E(\vec{s})$, the new state is accepted. However, if the energy is increased, we accept the flip with probability $\exp\left(-\beta \left(E(\pvec{s}') - E(\vec{s})\right)\right)$ where $\beta$ is the inverse temperature at that time. After each sweep, the inverse temperature is increased. We use $\beta_{\mathrm{init}} = 0.1$ and $\beta_{\mathrm{fin}}=10$ with a linear annealing schedule (at each sweep, $\beta$ is increased by $(\beta_{\mathrm{fin}}-\beta_{\mathrm{init}})/S$ where $S$ is the total number of sweeps performed). For the time complexity analysis (section~\ref{sec:results2}), we only seek to find a lower bound on the runtime of SA, and we thus select parameters to better estimate time to convergence at the cost of minimizing energy. The convergence time for each QUBO is computed after performing $S = 15000$ sweeps and $5$ runs of SA to obtain the lowest energy state for these schedule parameters. We then begin with $S=1$ and increment $S$ until we reach the previously found lowest energy, thus obtaining the number of sweeps required to converge. This process is carried out $5$ times to obtain $5$ samples for convergence times for each QUBO. We then use bootstrapping to compute mean convergence time and confidence intervals, generating $250$ pseudo-samples for every tracking event and sampling uniformly from the $5$ convergence time samples for each sub-QUBO.

Note that for large QUBOs in the $2000$-track data point, we weakened our criteria for convergence by accepting convergence if the found state was $\epsilon$-close to the converged low-energy state. However, this strictly reduces the number of sweeps required for the $2000$-tracks data point and still provides exponential scaling with respect to number of tracks. Weakening the lower bound through this heuristic still yielded exponential scaling, confirming the time complexity of SA.

In measuring convergence time, it is important to realize that the converged low-energy state might not be the global ground state. However, the time to ground-state solution must be bounded from below by the convergence time. Hence, given that convergence times grow exponentially with number of tracks as in figure~\ref{fig:sa_scale}, we expect time to solution to grow exponentially as well, indicating that the pre-processing methodology did not reduce the QUBO to a polynomial-time problem.

Additionally, to compare quantum and simulated annealing performance (section~\ref{sec:results_qasa}), we use 1000 reads and 1000 sweeps to ensure a lower energy is reached than in the timing analysis, since we wish to compare purity and efficiency rather than gain an exact estimate of time to convergence. No significant difference was observed when increasing the number of reads or sweeps in simulated annealing over the range of track densities we report.
 
\end{document}